\documentclass[aps,prd,10pt,tightenlines,amsmath,unsortedaddress,twocolumn,amssymb,10pt]{revtex4}

\usepackage{float}
\usepackage{graphicx}
\usepackage{dcolumn}
\usepackage{bm}
\usepackage{bbold}
\usepackage{enumerate}
\usepackage[utf8]{inputenc}
\usepackage[english]{babel}
\usepackage{amsthm}

\usepackage{xcolor}

\begin{document}

\title{ALP induced polarization effects on photons from galaxy clusters}

\author{Giorgio Galanti}
\email{gam.galanti@gmail.com}
\affiliation{INAF, Istituto di Astrofisica Spaziale e Fisica Cosmica di Milano, Via Alfonso Corti 12, I -- 20133 Milano, Italy}

\author{Marco Roncadelli}
\email{marcoroncadelli@gmail.com}
\affiliation{INFN, Sezione di Pavia, Via Agostino Bassi 6, I -- 27100 Pavia, Italy}
\affiliation{INAF, Osservatorio Astronomico di Brera, Via Emilio Bianchi 46, I -- 23807 Merate, Italy}

\author{Fabrizio Tavecchio}
\email{fabrizio.tavecchio@inaf.it}
\affiliation{INAF, Osservatorio Astronomico di Brera, Via Emilio Bianchi 46, I -- 23807 Merate, Italy}

\author{Enrico Costa}
\email{enrico.costa@inaf.it}
\affiliation{INAF, Istituto di Astrofisica e Planetologia Spaziali, Via del Fosso del Cavaliere 100, I -- 00133 Roma, Italy}

\date{\today}

\begin{abstract}

Many extensions of the Standard Model of particle physics and in particular superstring and superbrane theories predict the existence of axion-like particles (ALPs). ALPs are very elusive, extremely light and interact primarily with photons. In the presence of an external magnetic field two effects show up: (i) photon-ALP oscillations and (ii) a change of the photon polarization state. The astrophysical context represents the best opportunity to get indirect evidence for the ALP existence thanks to various effects that the photon-ALP interaction produces in the sky. Great attention has been paid so far to photon-ALP oscillations, since they modify the transparency of the crossed media at very high energies and so the final spectra of faraway sources exhibit a flux excess and a  characteristic oscillatory behavior. Two hints at the ALP existence have hitherto been discovered. But less interest has been attracted by the modification of the photon polarization. In this paper we address it in the X-ray and in the high energy (HE) bands. Specifically, we analyze the photon degree of linear polarization and the polarization angle induced by the photon-ALP interaction for photons generated in the central region of two galaxy clusters: Perseus and Coma. We find a substantial departure from conventional physics in both considered bands.  We conclude that the ALP-induced polarization effects are more likely detectable with the proposed missions like COSI (approved to launch), e-ASTROGAM and AMEGO in the HE range. Still, possible ALP-induced effects on photon polarization could also be detected by IXPE (already operative) and by the proposed eXTP, XL-Calibur, NGXP and XPP in the X-ray band.

\end{abstract}



\maketitle



\section{Introduction}

Astroparticle physics is nowadays a very exciting and promising research field. Just to quote two key-achievements, think of the detection of the right neutrinos flux from the Sun~\cite{neutrSun} and more recently the discovery of very-high-energy (VHE) neutrinos from blazars (a class of active galactic nuclei, AGN)~\cite{blazarNeutr}.  

More specifically, axion-like particles (ALPs, see e.g.~\cite{alp1,alp2}) are nowadays in the limelight since they are a generic prediction of many extensions of the Standard Model of particle physics including superstring and superbrane  theories~\cite{string1,string2,string3,string4,string5,axiverse,abk2010,cicoli2012,cisterna1,cisterna2}. Besides, they are among the best candidates for the dark matter~\cite{preskill,abbott,dine,sikivie1983,arias2012,jaekel}. Actually, ALPs are very similar to the axion -- the pseudo-Goldsone boson arising from the breakdown of the global Peccei-Quinn symmetry $\rm U(1)_{PQ}$ -- proposed to solve the strong CP problem (for a review, see~\cite{axionrev1,axionrev2,axionrev3,axionrev4}). But while axion mass and two-photon coupling are tightly related and axions couple to fermions and gluons, the ALP mass $m_a$ and the ALP two-photon coupling $g_{a\gamma\gamma}$ are {\it independent} parameters, and other possible ALP interactions are subdominant. In the presence of an external magnetic field two effects arise (for a review, see~\cite{irastorzaredondo,gRew,grRew}). 

\begin{enumerate}[(i)]
\item Photon-ALP oscillations~\cite{mpz,raffeltstodolsky} -- which are analogous to the oscillations of massive neutrinos of different flavors. 
\item The change of the polarization state of a photon beam~\cite{mpz,raffeltstodolsky}.
\end{enumerate} 

Effect (i) has in turn two main implications. One consists of an enhanced photon transparency of the sky at VHE, say above $100 - 1000 \, {\rm GeV}$ (see~\cite{drm2007,dgr2011}). The other amounts to an oscillatory behavior in the observed spectra of VHE blazars~\cite{dmr2008,wb2012,gr2013,grExt,gtre2019,gtl2020}. Very remarkably, two {\it hints} at the existence of an ALP have been found. The first one is that ALPs 
naturally explain why flat spectrum radio quasars (a class of blazars) can emit photons with energies up to $400 \, \rm GeV$ as observed -- without invoking {\it ad hoc} solutions -- while conventional physics prevents any photon emission above 30 GeV~\cite{trgb2012}. The second is that ALPs solve the anomalous redshift-dependence of the spectra of BL Lacs (another class of blazars)~\cite{grdb}. The gamma-ray burst GRB 221009A detected at $18 \, \rm TeV$ by LHAASO~\cite{LHAASO} or even at $251 \, \rm TeV$ by Carpet-2~\cite{carpet} would represent a firm indication for the ALP existence with the properties employed in the previous two hints~\cite{grtGRB}.

Effect (ii) has been already studied but with less interest: ALP-induced implications for the polarization of photons from gamma-ray bursts have been analyzed 
in~\cite{bassan} and the polarization of photons produced in the central region of a galaxy cluster or in the blazar jet in the presence of ALPs has been considered 
in~\cite{galantiPol}. Photon-ALP conversion effects on photon polarization from other astrophysical sources have been addressed  in~\cite{ALPpol1,ALPpol2,ALPpol3,ALPpol4,ALPpol5,day}. In addition, photon-ALP interaction has recently been discovered as a means to measure {\it emitted} photon polarization~\cite{galantiTheo}. All these studies are quite instrumental to find new hints at the ALP existence and they are timely, since new observatories measuring photon polarization have been launched or proposed in the X-ray band like IXPE~\cite{ixpe}, eXTP~\cite{extp}, XL-Calibur~\cite{xcalibur}, NGXP~\cite{ngxp} and XPP~\cite{xpp}, and in the high-energy (HE) band such as COSI~\cite{cosi}, e-ASTROGAM~\cite{eastrogam1,eastrogam2} and AMEGO~\cite{amego}.

In this paper we compute the photon survival probability in the presence of ALPs  $P_{\gamma \to \gamma}$, the related degree of linear polarization $\Pi_L$ and the polarization angle $\chi$ of a photon-ALP beam produced in the central region of two iconic galaxy clusters: Perseus and Coma. We follow the analysis developed in~\cite{galantiPol}. We use parameters within physically consistent bounds concerning both the astrophysical context and the photon-ALP system. Moreover, we investigate the propagation of the photon-ALP beam in all the crossed regions (galaxy cluster, extragalactic space, Milky Way) by using the state-of-the-art knowledge. In the considered scenarios ALPs manifestly induce photon polarization effects that are expected to be detectable by current and planned satellite missions~\cite{ixpe,extp,xcalibur,ngxp,xpp,cosi,eastrogam1,eastrogam2,amego}. We conclude that the HE range is the most promising window to search for ALP-induced polarization effects in galaxy clusters, but we cannot exclude the X-ray band.

The paper is organized as follows. In Sec. II we recall the main properties of ALPs and of the photon-ALP system, stressing polarization effects. In Sec. III we review the main properties of the astrophysical environments crossed by the photon-ALP beam, in Sec. IV we present our results about the Perseus and Coma clusters, while in Sec. V we draw our conclusions.

\section{Axion-like particles and polarization}

ALPs denoted by $a$ are very light neutral pseudo-scalar bosons interacting primarily with two photons and described by the Lagrangian
\begin{eqnarray}
&\displaystyle {\cal L}_{\rm ALP} =  \frac{1}{2} \, \partial^{\mu} a \, \partial_{\mu} a - \frac{1}{2} \, m_a^2 \, a^2 - \, \frac{1}{4 } g_{a\gamma\gamma} \, F_{\mu\nu} \tilde{F}^{\mu\nu} a \nonumber \\
&\displaystyle = \frac{1}{2} \, \partial^{\mu} a \, \partial_{\mu} a - \frac{1}{2} \, m_a^2 \, a^2 + g_{a\gamma\gamma} \, {\bf E} \cdot {\bf B}~a~,
\label{lagr}
\end{eqnarray}
where ${\bf E}$ and ${\bf B}$ are the electric and magnetic components of the electromagnetic tensor $F_{\mu\nu}$ with $\tilde{F}^{\mu\nu}$ representing its dual. While many limits on the photon-ALP coupling $g_{a \gamma \gamma}$ and ALP mass $m_a$ are present in the literature~\cite{cast,straniero,fermi2016,payez2015,berg,conlonLim,meyer2020,limFabian,limJulia,limKripp,limRey2,mwd}, the only firm bound is represented by $g_{a \gamma \gamma} < 0.66 \times 10^{- 10} \, {\rm GeV}^{- 1}$ for $m_a < 0.02 \, {\rm eV}$ at the $2 \sigma$ level from no detection of ALPs from the Sun by CAST~\cite{cast}.

In the VHE range we should consider also the Heisenberg-Euler-Weisskopf (HEW) effective Lagrangian
\begin{equation}
\label{HEW}
{\cal L}_{\rm HEW} = \frac{2 \alpha^2}{45 m_e^4} \, \left[ \left({\bf E}^2 - {\bf B}^2 \right)^2 + 7 \left({\bf E} \cdot {\bf B} \right)^2 \right]~,
\end{equation}
which takes into account the photon one-loop vacuum polarization effects with $\alpha$ and $m_e$ denoting the fine-structure constant and the electron mass, respectively~\cite{hew1, hew2, hew3}. However, the effects of ${\cal L}_{\rm HEW}$ turn out to be totally irrelevant in the X-ray and HE band for the systems we are dealing with~\cite{grSM,grExt}. The same conclusion remains true concerning the photon dispersion on the cosmic microwave background (CMB)~\cite{raffelt2015} (see also~\cite{grSM,grExt}).

We consider in the following a photon-ALP beam ($\bf E$ in Eq.~(\ref{lagr}) pertains to a propagating photon) of energy $E$, propagating from a galaxy cluster towards us along the $y$-direction and crossing several magnetized media (galaxy cluster, extragalactic space, Milky Way, see Sec. III for more details). Note that the presence of an external magnetic field is crucial in order for photon-ALP oscillations to take place so as to compensate the spin mismatch between photons and ALPs. Photon-ALP oscillations can show up since the propagation eigenstates differ from the interaction eigenstates for the off-diagonality of the mass matrix of the $\gamma - a$ system. Because of the structure of the interaction term in ${\cal L}_{\rm ALP}$, only the transverse component of $\bf B$ and denoted by ${\bf B}_T$ couples to $a$.

By successfully employing the short-wavelength approximation~\cite{raffeltstodolsky} -- since we have $E \gg m_a$ -- the propagation equation along the $y$-direction of a non-polarized photon-ALP beam arising from ${\cal L}_{\rm ALP}$ reads
\begin{equation}
\label{vneum}
i \frac{d \rho (y)}{d y} = \rho (y) \, {\cal M}^{\dag} ( E, y) - {\cal M} ( E, y) \, \rho (y)~,
\end{equation}
where ${\cal M} ( E, y)$ is the mixing matrix of the photon-ALP system and accounts for the photon-ALP interaction strength, the ALP and effective photon mass, the magnetization and absorption properties of the crossed medium (for more details see~\cite{grSM}). In Eq.~(\ref{vneum}) $\rho(y)$ represents the polarization density matrix of the photon-ALP system, which reads
\begin{equation}
\label{densmat}
\rho (y) = \left(\begin{array}{c}A_x (y) \\ A_z (y) \\ a (y)
\end{array}\right)
\otimes \left(\begin{array}{c}A_x (y) \  A_z (y) \ a (y) \end{array}\right)^{*}~,
\end{equation}
where $A_x(y)$ and $A_z(y)$ are the photon linear polarization amplitudes along the $x$ and $z$ axis, respectively and $a(y)$ is the ALP amplitude. The solutions of Eq.~(\ref{vneum}) can be expressed in terms of the {\it transfer matrix} of the photon-ALP system ${\cal U} \bigl( E; y, y_0 \bigr)$ as
\begin{equation}
\label{unptrmatr}
\rho ( y ) = {\cal U} \bigl(E; y, y_0 \bigr) \, \rho_0 \, {\cal U}^{\dag} \bigl(E; y, y_0 \bigr)~,
\end{equation}
where $\rho_0$ is the density matrix at position $y_0$. Now, the probability that a photon-ALP beam initially in the state $\rho_0$ at position $y_0$ will be found at position $y$ in the state $\rho$ reads
\begin{equation}
\label{unpprob}
P_{\rho_0 \to \rho} (E,y) = {\rm Tr} \Bigl[\rho \, {\cal U} (E; y, y_0) \, \rho_0 \, {\cal U}^{\dag} (E; y, y_0) \Bigr]~,
\end{equation}
with ${\rm Tr} \, \rho_0 = {\rm Tr} \, \rho =1$~\cite{dgr2011}.

Equation~(\ref{densmat}) can be specialized to describe pure photon states in the $x$ and $z$ direction as
\begin{equation}
\label{densphot}
{\rho}_x = \left(
\begin{array}{ccc}
1 & 0 & 0 \\
0 & 0 & 0 \\
0 & 0 & 0 \\
\end{array}
\right)~, \,\,\,\,\,\,\,\,
{\rho}_z = \left(
\begin{array}{ccc}
0 & 0 & 0 \\
0 & 1 & 0 \\
0 & 0 & 0 \\
\end{array}
\right)~,
\end{equation}
respectively, and the ALP state as
\begin{equation}
\label{densa}
{\rho}_a = \left(
\begin{array}{ccc}
0 & 0 & 0 \\
0 & 0 & 0 \\
0 & 0 & 1 \\
\end{array}
\right)~,
\end{equation}
while unpolarized photons are represented by
\begin{equation}
\label{densunpol}
{\rho}_{\rm unpol} = \frac{1}{2} \left(
\begin{array}{ccc}
1 & 0 & 0 \\
0 & 1 & 0 \\
0 & 0 & 0 \\
\end{array}
\right)~.
\end{equation}
Partially polarized photons can be described by a polarization density matrix with an intermediate functional expression between Eqs.~(\ref{densphot}) and Eq.~(\ref{densunpol}). 

We can now write down the photonic part of the polarization density matrix in Eq.~(\ref{densmat}) in terms of the Stokes parameters as~\cite{poltheor1}
\begin{equation}
\label{stokes}
{\rho}_{\gamma} = \frac{1}{2} \left(
\begin{array}{cc}
I+Q & U-iV \\
U+iV & I-Q \\
\end{array}
\right)~,
\end{equation}
while the definition of the photon degree of {\it linear polarization} $\Pi_L$ and of the {\it polarization angle} $\chi$ read~\cite{poltheor2}
\begin{equation}
\label{PiL}
\Pi_L \equiv \frac{(Q^2+U^2)^{1/2}}{I}~,
\end{equation}
\begin{equation}
\label{chiPol}
\chi \equiv \frac{1}{2}{\rm atan}\left(\frac{U}{Q}\right)~,
\end{equation}
which in terms of the photon polarizaton density matrix elements $\rho_{ij}$ with $i,j=1,2$ can be expressed as
\begin{equation}
\label{PiL2}
\Pi_L = \frac{\left[ (\rho_{11}-\rho_{22})^2+(\rho_{12}+\rho_{21})^2\right]^{1/2}}{\rho_{11}+\rho_{22}}~,
\end{equation}
and
\begin{equation}
\label{chiPol2}
\chi = \frac{1}{2}{\rm atan}\left(\frac{\rho_{12}+\rho_{21}}{\rho_{11}-\rho_{22}}\right)~,
\end{equation}
respectively.

In the absence of photon absorption, there exists a strict relationship between the {\it emitted} photon degree of linear polarization $\Pi_{L,0}$ and the photon survival probability $P_{\gamma \to \gamma}$ as the theorems stated and demonstrated in~\cite{galantiTheo} show. For the specific physical cases in question, since $\Pi_{L,0} = 0$ (more about this in Sec. III), the theorems in~\cite{galantiTheo} ensure that $P_{\gamma \to \gamma} \geq 1/2$, as the following figures show.

\section{Photon-ALP beam propagation} 

We now cursorily describe the astrophysical media crossed by the photon-ALP beam (galaxy cluster, extragalactic space, Milky Way) by stressing their fundamental properties which are important for the photon-ALP conversion. We develop a strategy similar to the analysis performed in~\cite{galantiPol} and we refer the reader to the publications cited below about the specific topic for more details. As benchmark cases, the photon-ALP system is assumed to possess $g_{a\gamma\gamma}= 0.5 \times 10^{-11} \, \rm GeV^{-1}$, while for the ALP mass we take: (i) $m_a \lesssim 10^{-14} \, \rm eV$ and (ii) $m_a = 10^{-10} \, \rm eV$, which allow us to satisfy the most solid bound present in the literature~\cite{cast}.

\subsection{Galaxy cluster}

Galaxy clusters are the largest gravitationally bound structures in the Universe and consist of 30 to about 1000 galaxies, with a total mass in the range $(10^{14}-10^{15}) \, M_{\odot}$. Clusters can be divided into three main classes: (i) regular clusters, (ii) intermediate clusters, and (iii) irregular clusters. Many properties are associated with the class to which the cluster belongs, such as shape, symmetry, concentration, density profile, galactic content~\cite{GalCluClass}. We will concentrate in this paper on regular clusters since their approximate spherical symmetry and density profile provide us the possibility to model them in a rather accurate way. In particular, we are concerned with the strength and morphology of the cluster magnetic field ${\bf B}^{\rm clu}$, which is linked to the cluster electron number density $n_e^{\rm clu}$. And a precise modeling of ${\bf B}^{\rm clu}$ is crucial in order to estimate in an accurate fashion the photon-ALP conversion in the cluster.

It is nowadays well established from Faraday rotation measurements and synchrotron radio emission that $B^{\rm clu} = {\cal O} (1-10) \, \mu{\rm G}$~\cite{cluB1,cluB2} and that it possesses an isotropic Gaussian turbulent nature with a Kolmogorov-type turbulence power spectrum $M(k)\propto k^q$, with $k$ as the wave number in the range $[k_L,k_H]$ and index $q=-11/3$~\cite{cluFeretti}. The lower and upper limits of the latter interval $k_L = 2\pi/\Lambda_{\rm max}$ and $k_H = 2\pi/\Lambda_{\rm min}$ are the minimal and maximal turbulence scales, respectively. Thus, the behavior of ${\bf B}^{\rm clu}$ is modeled as~\cite{cluFeretti,clu2}
\begin{equation}
\label{eq1}
B^{\rm clu}(y)={\cal B} \left( B_0^{\rm clu},k,q,y \right) \left( \frac{n_e^{\rm clu}(y)}{n_{e,0}^{\rm clu}} \right)^{\eta_{\rm clu}}~,
\end{equation}
where ${\cal B}$ is the spectral function describing the Kolmogorov-type turbulence of the cluster magnetic field (for more details, see e.g.~\cite{meyerKolm}), $B_0^{\rm clu}$ and $n_{e,0}^{\rm clu}$ are the central cluster magnetic field strength and the central electron number density, respectively, while $\eta_{\rm clu}$ is a cluster parameter.

Concerning the profile of $n_e^{\rm clu}$, various models exist in the literature, such as the single $\beta$ model, the double $\beta$ model or a modified version of them~\cite{cluValues}. Usually, the single $\beta$ model is used to describe non-cool-core (nCC) clusters, while the double $\beta$ model fits better cool-core (CC) clusters. However, this distinction is not sharp and the single $\beta$ model is often employed for both CC and nCC clusters~\cite{cluValues}. Yet, we do not need to choose a theoretical model in the following as we will consider two specific clusters -- Perseus and Coma -- for which specific  models exist in the literature. We describe the employed models of $n_e^{\rm clu}$ for Perseus and Coma in the dedicated subsections of Sec. IV below.

Photons are produced in the cluster central region via different processes, such as thermal Bremsstrahlung showing up in the X-ray band~\cite{mitchell1979}, and likely synchrotron radiation in the cluster turbulent magnetic field of electrons produced by the cascade of VHE photons, inverse Compton scattering and neutral pion decay generated in several ways in the HE band (see e.g.~\cite{cluGammaEm1,cluGammaEm2,cluGammaEm3,cluGammaEm4}). In all these cases the emitted photons are effectively unpolarized~\cite{polarRev,FermiPol} -- in the case of synchrotron radiation because of the turbulent nature of ${\bf B}^{\rm clu}$ (see also note~\cite{noteCluPol}). We shall keep in mind that, while cluster emission in the X-ray band is corroborated by many observations, photon production in the HE range is less solid from an observational point of view for the lack of substantial detection.

By studying the propagation of the photon-ALP beam starting from the cluster central region out to its virial radius we compute the transfer matrix ${\cal U}_{\rm clu}$ of the photon-ALP system in the cluster.

\subsection{Extragalactic space}

Because we are concerned with the photon-ALP beam propagation in the X-ray and in HE band for energies $E_0 \leq 10 \, \rm GeV$, photon absorption by the extragalactic background light (EBL) is totally negligible ~\cite{franceschinirodighiero,dgr2013,gptr}.

The strength and morphology of the extragalactic magnetic field ${\bf B}_{\rm ext}$ is nowadays poorly known: current bounds restrict $B_{\rm ext}$ to the range $10^{- 7} \, {\rm nG} \leq {B}_{\rm ext} \leq 1.7 \, {\rm nG}$ on the scale of ${\cal O} (1) \, {\rm Mpc}$~\cite{neronov2010,durrerneronov,pshirkov2016}. The shape of ${\bf B}_{\rm ext}$ is modeled by means of a {\it domain-like} structure: in each domain of size $L_{\rm dom}^{\rm ext}$ which is equal to the magnetic field coherence length, ${\bf B}_{\rm ext}$ possesses a constant strength and the same orientation, which changes randomly and {\it discontinuously} from one domain to the next~\cite{kronberg1994,grassorubinstein}. Concerning the strength and coherence of ${\bf B}_{\rm ext}$, quite high values are predicted by outflows from primeval galaxies with $B_{\rm ext} = {\cal O}(1) \, \rm nG$ for $L_{\rm dom}^{\rm ext} = {\cal O}(1) \, \rm Mpc$~\cite{reessetti,hoyle,kronberg1999,furlanettoloeb}. Thus, we take $B_{\rm ext}=1 \, \rm nG$ and $L_{\rm dom}^{\rm ext}$ randomly varying with a power-law distribution function $\propto (L_{\rm dom}^{\rm ext})^{-1.2}$ in the range $(0.2 -10) \, \rm Mpc$ and with $\langle L_{\rm dom}^{\rm ext} \rangle = 2 \, \rm Mpc$.

We want to stress that the simple above-described domain-like model is not always appropriate to describe ${\bf B}_{\rm ext}$: when the oscillation length $l_{\rm osc}$ of the photon-ALP beam turns out to be smaller than $L_{\rm dom}^{\rm ext}$, the system becomes sensitive to the ${\bf B}_{\rm ext}$ substructure, so that the standard discontinuous model produces unphysical results. Therefore, we employ a new model developed in~\cite{grSM} that preserves the features of the domain-like model but continuously connects the ${\bf B}_{\rm ext}$ components crossing the boundary between any two adjacent domains. As a consequence, the photon-ALP beam propagation can still be {\it analytically} computed, leading to physically consistent results despite an increase of computational complexity (for more details see~\cite{grSM}).

By means of the procedure developed in~\cite{grExt,grSM}, we can calculate the transfer matrix of the photon-ALP system in the extragalactic space ${\cal U}_{\rm ext}$.

\subsection{Milky Way}

The study of the photon-ALP conversion in the Milky Way is facilitated by the existence of quite accurate maps of both the electron number density $n_e^{\rm MW}$ and the magnetic field ${\bf B}_{\rm MW}$. Concerning $n_e^{\rm MW}$ we use the model developed in~\cite{ymw2017}. The structure of ${\bf B}_{\rm MW}$ is quite complex, presenting both a regular and a turbulent part. The regular component of ${\bf B}_{\rm MW}$ with strength of ${\cal O}(1) \, \mu{\rm G}$ is responsible for the dominant effect on the photon-ALP beam propagation. Instead, the effect produced by the turbulent part is negligible, since the photon-ALP beam oscillation length is much larger than the coherence length of the turbulent component of ${\bf B}_{\rm MW}$.

Yet, we accurately model both the regular and the turbulent part of ${\bf B}_{\rm MW}$ by employing the model by Jansson and Farrar~\cite{jansonfarrar1,jansonfarrar2,BMWturb}, which includes a disk and a halo component, both parallel to the Galactic plane, and poloidal `X-shaped' component at the galactic center. In the literature there exists also the ${\bf B}_{\rm MW}$ model of Pshirkov {\it et al.}~\cite{pshirkovMF2011}, but it fails to accurately account for the Galactic halo component. Therefore, since we have checked that our results are not qualitatively modified by employing the model by Pshirkov {\it et al.}~\cite{pshirkovMF2011}, we use the model by Jansson and Farrar~\cite{jansonfarrar1,jansonfarrar2,BMWturb} to evaluate the transfer matrix ${\cal U}_{\rm MW}$ of the photon-ALP system inside the Milky Way by following the strategy developed in~\cite{gtre2019}.

\subsection{Overall photon-ALP beam propagation}
We can now calculate the total transfer matrix $\cal U$ of the photon-ALP system by multiplying ${\cal U}_{\rm clu}$, ${\cal U}_{\rm ext}$ and ${\cal U}_{\rm MW}$ in the correct order as
\begin{equation} 
\label{Utot}
{\cal U}={\cal U}_{\rm MW}\,{\cal U}_{\rm ext}\,{\cal U}_{\rm clu}~.
\end{equation}
By specializing Eq.~(\ref{unpprob}), the photon survival probability of photons produced in the cluster central zone and oscillating into ALPs up to us, reads 
\begin{equation} 
\label{probSurvFinal}
P_{\gamma \to \gamma}  = \sum_{i = x,z} {\rm Tr} \left[\rho_i \, {\cal U} \, \rho_{\rm in} \, {\cal U}^{\dagger} \right]~,
\end{equation}
where $\rho_x$ and $\rho_z$ are expressed by Eqs.~(\ref{densphot}), while $\rho_{\rm in}$ represents the beam initial polarization density matrix. As we have discussed in Sec. III.A, since photons are expected to be produced unpolarized in the cluster central zone, $\rho_{\rm in}$ reads from Eq.~(\ref{densunpol}) and in particular $\rho_{\rm in} \equiv {\rho}_{\rm unpol}$. The final photon degree of linear polarization $\Pi_L$ and the polarization angle $\chi$ are obtained from Eq.~(\ref{PiL2}) and Eq.~(\ref{chiPol2}), respectively by recalling Eq.~(\ref{unptrmatr}) with $\rho_0 \equiv \rho_{\rm in} \equiv {\rho}_{\rm unpol}$.

\section{Results}

We are now in a position to study the final photon survival probability $P_{\gamma \to \gamma}$, the corresponding photon degree of linear polarization $\Pi_L$ and the polarization angle $\chi$ of the photon-ALP beam, after propagation in the regions considered in Sec. III (galaxy cluster, extragalactic space and Milky Way). We address two real cases, namely the photons produced in the central region of the Perseus and Coma clusters.          

For the parameters of the photon-ALP system we take $g_{a\gamma\gamma}=0.5 \times 10^{-11} \, \rm GeV^{-1}$ and two values for the ALP mass: (i) $m_a \lesssim 10^{-14} \, \rm eV$, (ii) $m_a = 10^{-10} \, \rm eV$. The ALP mass term is smaller than the plasma frequency for $m_a \lesssim 10^{-14} \, \rm eV$, while the opposite is true in the case $m_a = 10^{-10} \, \rm eV$.

Besides $P_{\gamma \to \gamma}$, $\Pi_L$ and $\chi$, we compute the probability density function $f_{\Pi}$ associated with $\Pi_L$ when many realizations of the photon-ALP beam propagation are considered. We deal with photons produced at redshift $z$ with energy $E$ and observed energy $E_0=E/(1+z)$ in the two ranges: (i) UV-X-ray band ($10^{-3}\, {\rm keV} -10^2 \, \rm keV$), (ii) HE band $(10^{-1} \, {\rm MeV} -10^4 \, \rm MeV)$.

In order to see whether the ALP-induced features of $\Pi_L$ and $\chi$ can be detected by real observatories, we have binned our final results with the appropriate energy resolution derived by current instrument capabilities (more about this below). But then it is mandatory to investigate whether or not the ALP-induced features are washed out by the binning procedure. Therefore, we bin our theoretical results about the Stokes parameters by computing the mean $\mu$ and the variance $\sigma$ of our energy points by assuming the realistic energy resolution of current observatories. In particular, we perform the following procedure: once the energy bin width is selected based on the instrument capabilities, we simply collect all the theoretical data within the bin and we directly calculate both $\mu$ and $\sigma$ without any weighting for the heuristic nature of our calculation. This procedure is performed for each bin. We next evaluate the error bars for the derived quantities $\Pi_L$ and $\chi$ by following the standard theory of the propagation of uncertainty (see~\cite{VarStokes} for a very detailed analysis about the realistic variance of the Stokes parameters).

Actually, since in the X-ray band the polarization measurements are more involved than the spectral ones, the energy resolution ought to be worsened by a factor of $4 - 5$. By considering the energy resolution of spectrum-measuring observatories in the X-ray band (see e.g. the energy resolution of SWIFT~\cite{swift}), we expect that the energy resolution for polarization studies should likely be of $15 - 20$ bins per decade in the X-ray band. Instead, in the HE range a similar resolution for spectral and polarimetric measurements is expected since they derive from the same data. By considering the energy resolution of future HE observatories~\cite{eastrogam1,eastrogam2,amego}, we conservatively assume a lower resolution of $8 - 10$ bins per decade. In the following, a signal of $\Pi_L >0$ is assumed as detectable if the $1 \sigma$ lower bounds of the data points are above zero (see also~\cite{noteDetectability}). Note that also $\Pi_L = 0$ represents a detection, namely of unpolarized photons.

\subsection{Perseus cluster}

\begin{figure}
\centering
\includegraphics[width=0.5\textwidth]{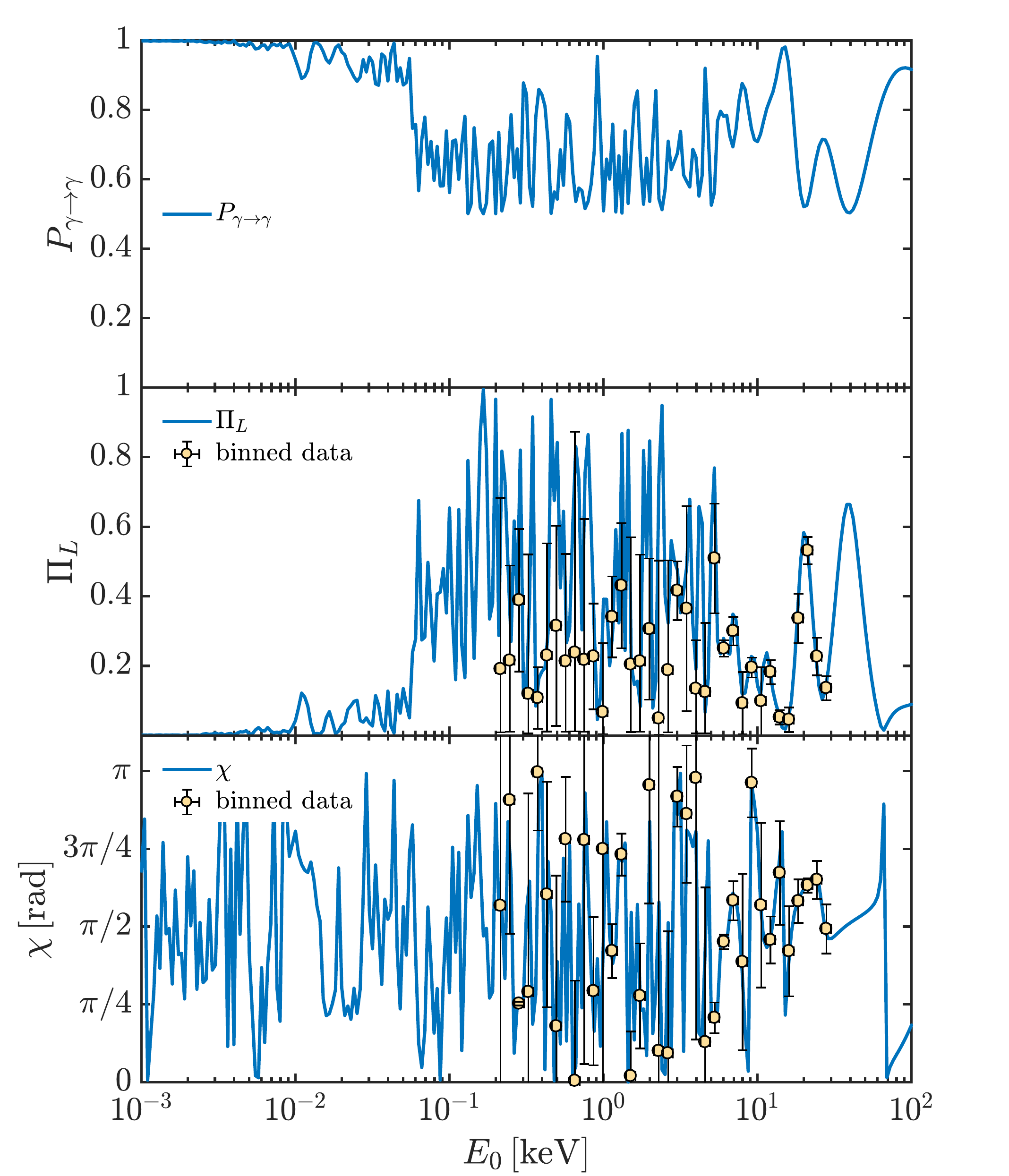}
\caption{\label{perseusAllPolKeV} Perseus cluster: photon survival probability $P_{\gamma \to \gamma }$ (upper panel), corresponding final degree of linear polarization $\Pi_L$ (central panel) and final polarization angle $\chi$ (lower panel) in the energy range $(10^{-3}-10^2) \, {\rm keV}$. We take $g_{a\gamma\gamma}=0.5 \times 10^{-11} \, \rm GeV^{-1}$, $m_a \lesssim 10^{-14} \, \rm eV$. The initial degree of linear polarization is $\Pi_{L,0}=0$.}
\end{figure}

\begin{figure}
\centering
\includegraphics[width=0.5\textwidth]{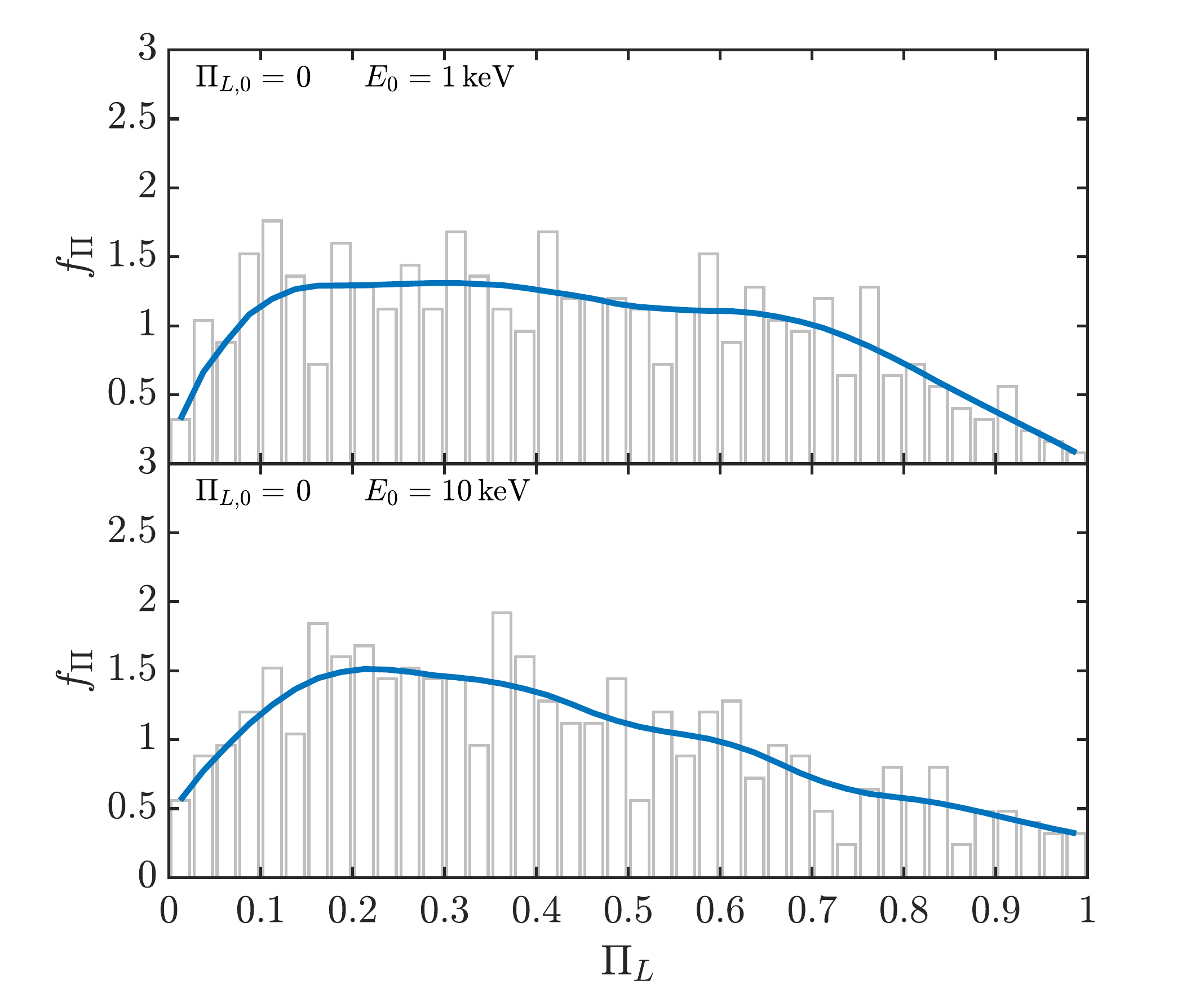}
\caption{\label{perseusDensKeV} Perseus cluster: probability density function $f_{\Pi}$ arising from the plotted histogram for the final degree of linear polarization $\Pi_L$ at $1 \, \rm keV$ (upper panel) and $10 \, \rm keV$ (lower panel) by considering the system in Fig.~\ref{perseusAllPolKeV}. The initial photon degree of linear polarization is $\Pi_{L,0}=0$.}
\end{figure}

\begin{figure}
\centering
\includegraphics[width=0.5\textwidth]{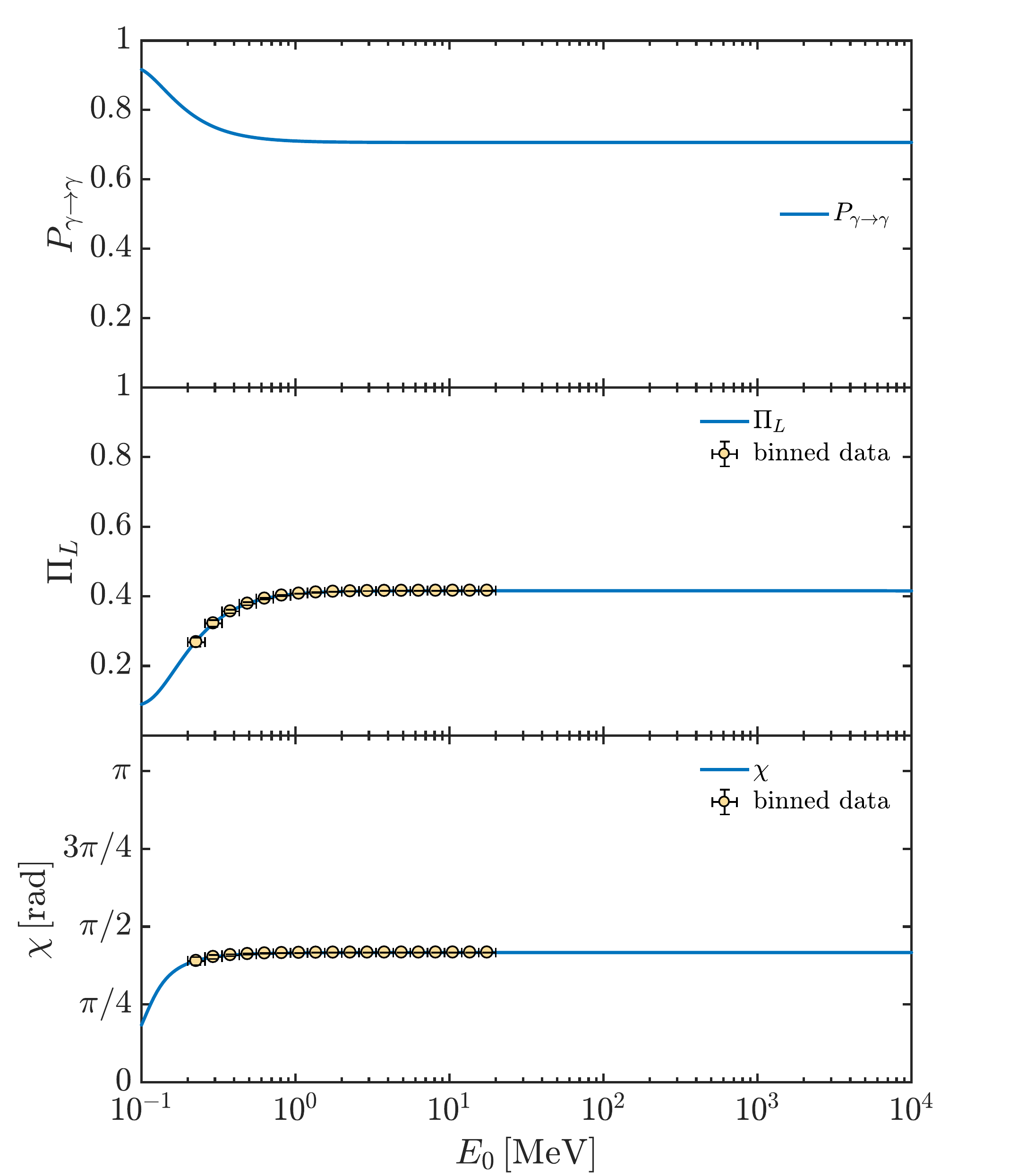}
\caption{\label{perseusAllPolMeV} Perseus cluster: same as Fig.~\ref{perseusAllPolKeV} but in the energy range $(10^{-1} -10^4) \, {\rm MeV}$. We take $g_{a\gamma\gamma}=0.5 \times 10^{-11} \, \rm GeV^{-1}$, $m_a \lesssim 10^{-14} \, \rm eV$. The initial degree of linear polarization is $\Pi_{L,0}=0$.}
\end{figure}

\begin{figure}
\centering
\includegraphics[width=0.5\textwidth]{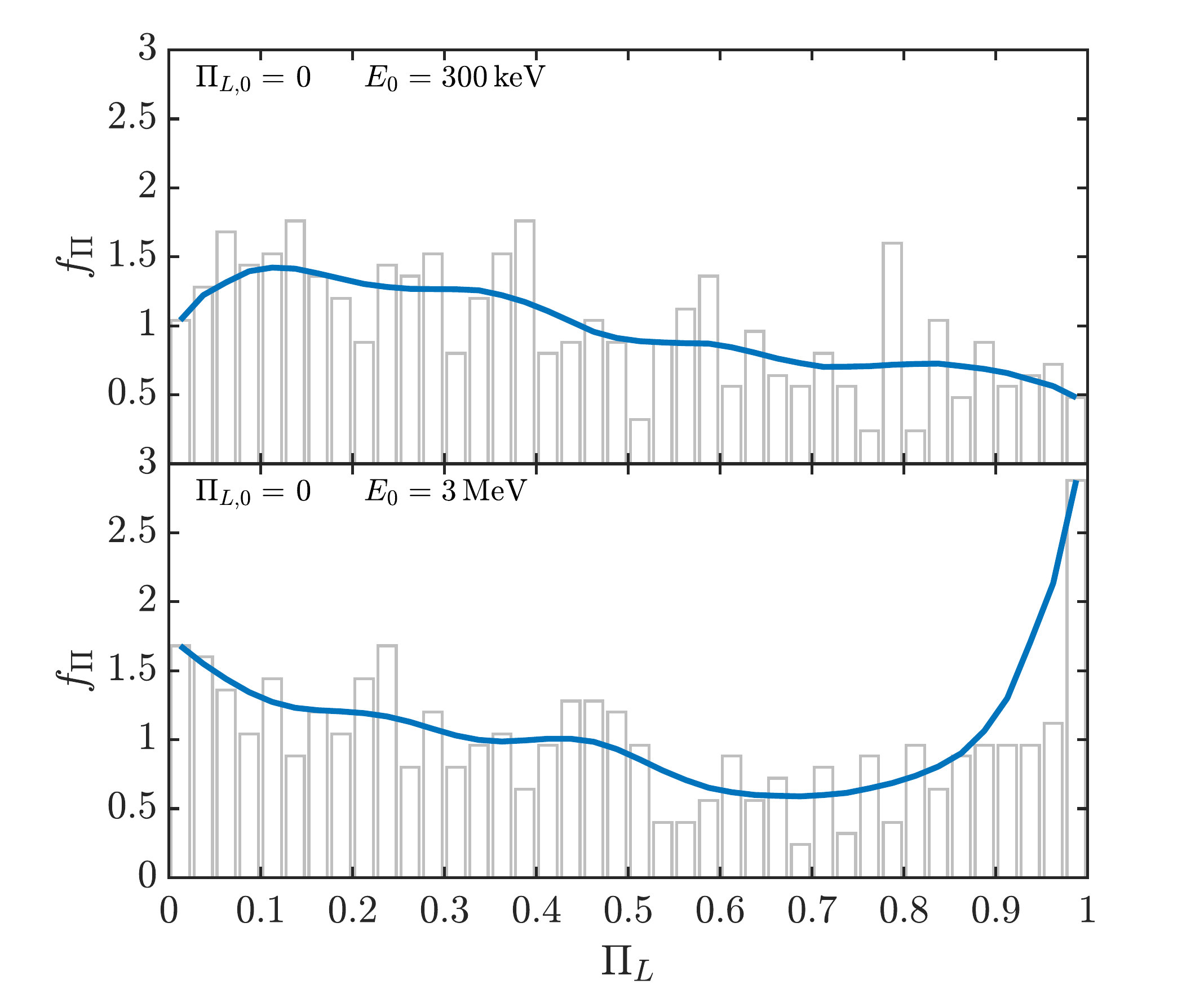}
\caption{\label{perseusDensMeV} Perseus cluster: same as Fig.~\ref{perseusDensKeV} but for the energies $300 \, \rm keV$ (upper panel) and $3 \, \rm MeV$ (lower panel) by considering the system in Fig.~\ref{perseusAllPolMeV}. The initial photon degree of linear polarization is $\Pi_{L,0}=0$.}
\end{figure}

\begin{figure}
\centering
\includegraphics[width=0.5\textwidth]{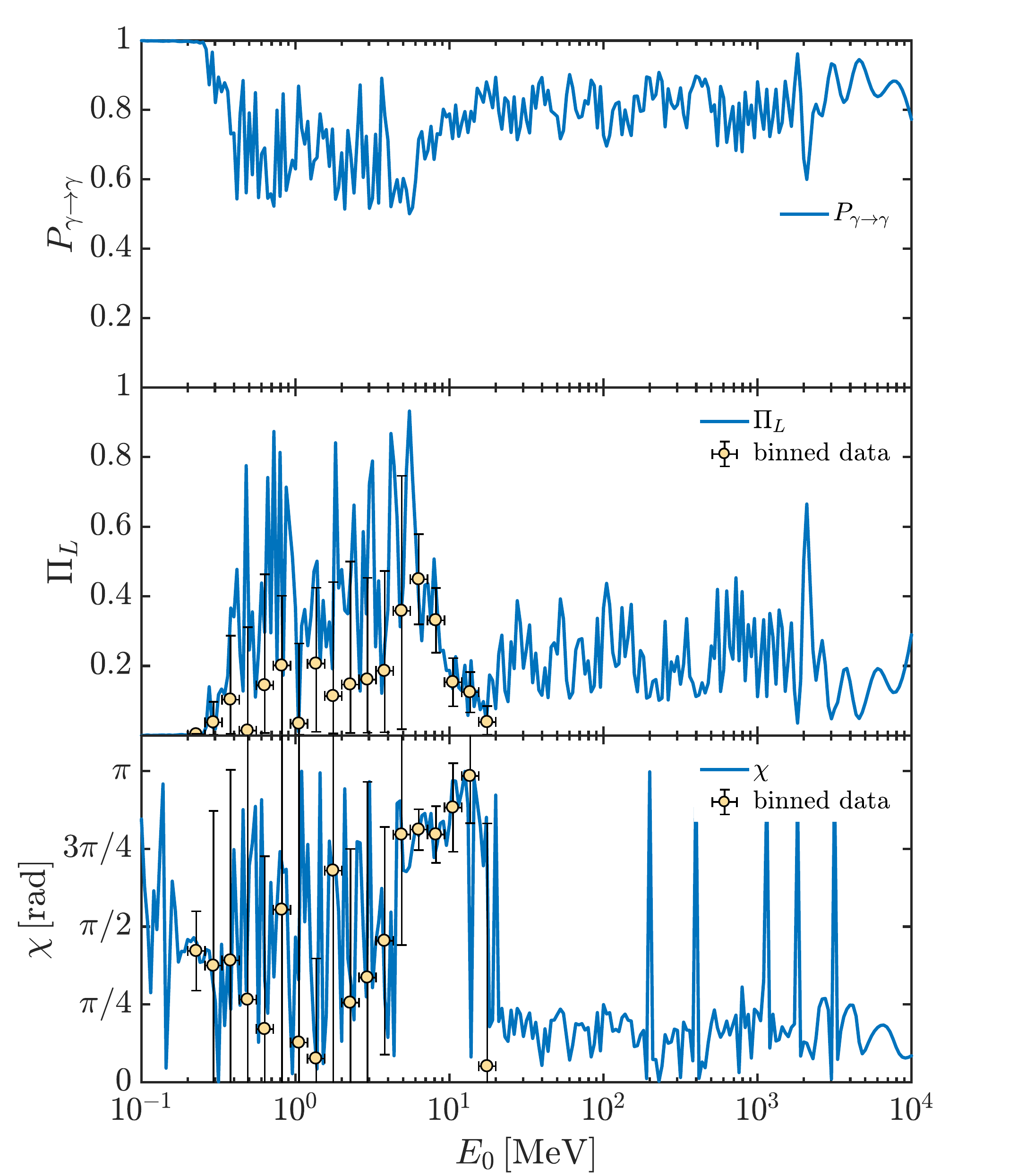}
\caption{\label{perseusAllPolMeV-10} Perseus cluster: same as Fig.~\ref{perseusAllPolMeV}. We take $g_{a\gamma\gamma}=0.5 \times 10^{-11} \, \rm GeV^{-1}$, $m_a = 10^{-10} \, \rm eV$. The initial degree of linear polarization is $\Pi_{L,0}=0$.}
\end{figure}

\begin{figure}
\centering
\includegraphics[width=0.5\textwidth]{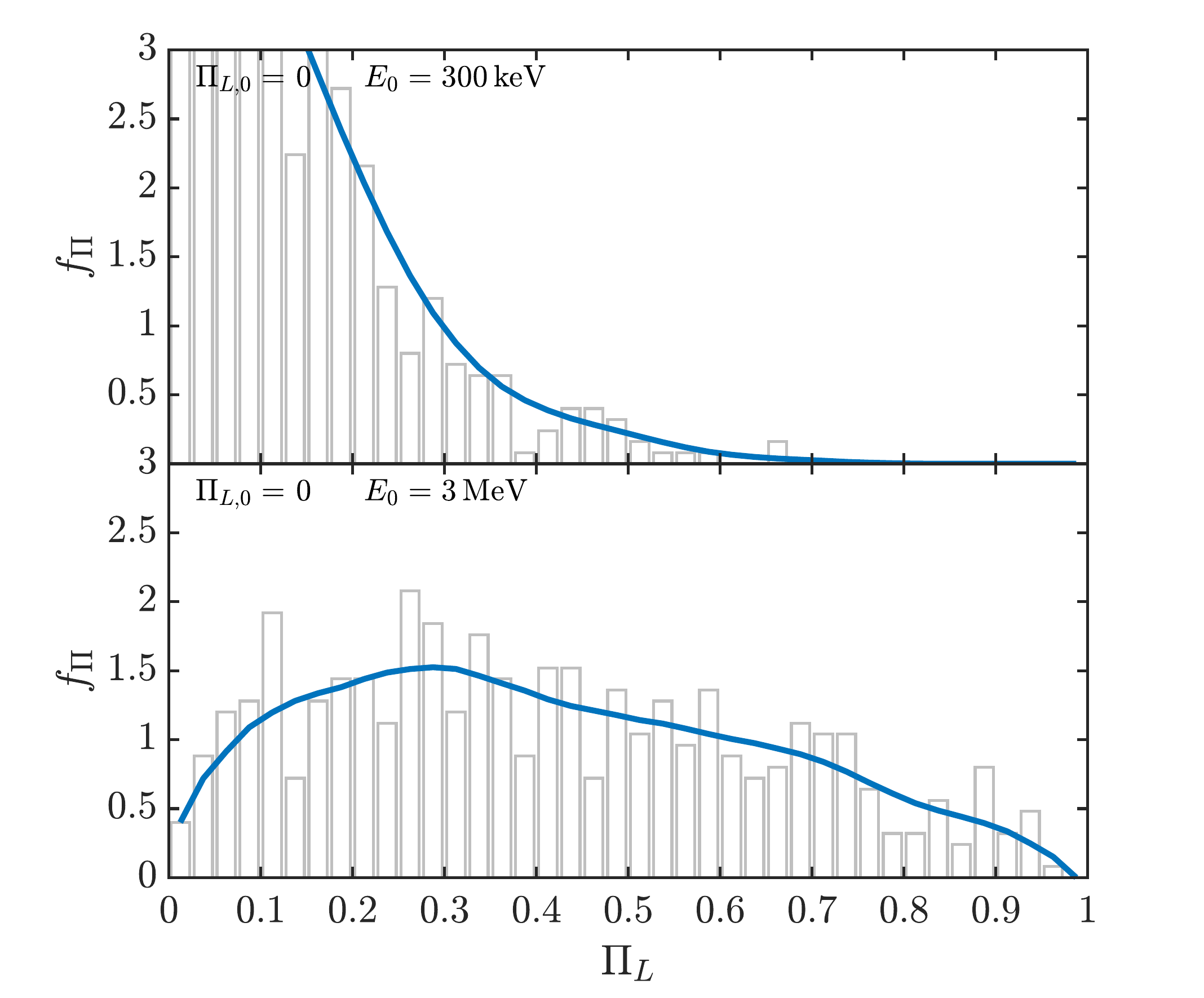}
\caption{\label{perseusDensMeV-10} Perseus cluster: same as Fig.~\ref{perseusDensMeV} by considering the system in Fig.~\ref{perseusAllPolMeV-10}. The initial photon degree of linear polarization is $\Pi_{L,0}=0$.}
\end{figure}

Perseus is a beautiful example of CC rich regular cluster. It is located at redshift $z = 0.01756$ and represents the brightest cluster in the X-ray sky. We describe the 
Perseus electron number density $n_e^{\rm clu}$ with the model~\cite{nePerseus}
\begin{eqnarray}
\label{eq2p}
&\displaystyle n_e^{\rm clu}(y)=n_{e,01}^{\rm clu} \left( 1+\frac{y^2}{r_{\rm core,1}^2} \right)^{-\frac{3}{2}\beta_{\rm clu,1}}+ \nonumber \\
&\displaystyle n_{e,02}^{\rm clu} \left( 1+\frac{y^2}{r_{\rm core,2}^2} \right)^{-\frac{3}{2}\beta_{\rm clu,2}}~,
\end{eqnarray}
where $n_{e,01}^{\rm clu}=3.9 \times 10^{-2} \, \rm cm^{-3}$, $r_{\rm core,1}=80 \, \rm kpc$, $\beta_{\rm clu,1}=1.2$, $n_{e,02}^{\rm clu}=4.05 \times 10^{-3} \, \rm cm^{-3}$, $r_{\rm core,2}=280 \, \rm kpc$ and $\beta_{\rm clu,2}=0.58$.

As far as the Perseus magnetic field ${\bf B}^{\rm clu}$ is concerned, we use Eq.~(\ref{eq1}) with $B_0^{\rm clu} = 16 \, \rm \mu G$. This choice represents an average value between that derived in~\cite{BpersLow} of $(2-13) \, \rm \mu G$ and the maximal estimate of $25 \, \rm \mu G$ as reported in~\cite{BpersHigh}. The stochastic properties of the Perseus magnetic field are taken from the study of the Coma cluster~\cite{cluFeretti}. This assumption is justified since the two clusters possess similar stochastic properties. Thus, we consider $q=-11/3$, $\Lambda_{\rm min}= 2 \, \rm kpc$ and $\Lambda_{\rm max}= 34 \, \rm kpc$~\cite{cluFeretti}. Finally, we take the average value $\eta_{\rm clu}=0.5$ entering Eq.~(\ref{eq1}). As discussed in Sec. III.A, photons emitted in the central region of a galaxy cluster (see also note~\cite{noteNGC1275}) are unpolarized both in the X-ray and in the HE band: thus, we accordingly assume an initial degree of linear polarization $\Pi_{L,0}=0$.

We start with the case $m_a \lesssim 10^{-14} \, \rm eV$. Our results in the UV-X-ray band ($10^{-3} \, {\rm keV}-10^2 \, \rm keV$) are shown in Figs.~\ref{perseusAllPolKeV} and~\ref{perseusDensKeV}. We plot $P_{\gamma \to \gamma}$ in the top panel of Fig.~\ref{perseusAllPolKeV} and the corresponding final $\Pi_L$ and $\chi$ in the central and lower panel of Fig.~\ref{perseusAllPolKeV}, respectively. From Fig.~\ref{perseusAllPolKeV} we see that the photon-ALP interaction starts to become efficient for $E_0 \gtrsim 10^{-2} \, \rm keV$, where $P_{\gamma \to \gamma} \neq 1$ and the corresponding $\Pi_L$ begins to increase showing that $\Pi_L >0$. We note that $P_{\gamma \to \gamma}$ and the corresponding $\Pi_L$ show an energy oscillatory behavior because the photon-ALP system is in the weak-mixing regime, as a consequence of the non-negligible plasma term (see also~\cite{grSM}). As already noted in~\cite{galantiPol} for an analogous physical system, the weak-mixing regime extends for almost four energy decades ($10^{-2}\, {\rm keV} - 10^2 \, \rm keV$) because of the high variation of ${\bf B}^{\rm clu}$ and $n_e^{\rm clu}$ as expressed by Eqs.~(\ref{eq1}) and Eq.~(\ref{eq2p}), respectively, in addition to the properties of the other crossed media. The behavior of $\chi$ confirms an high energy dependence of the system for $E_0 \lesssim 10^2 \, \rm keV$. While a detection of $\Pi_L >0$ appears as prohibitive for $E_0 \lesssim 1 \, \rm keV$, as the binned data in Fig.~\ref{perseusAllPolKeV} show, we expect not only to be able to detect a possible signal of $\Pi_L > 0$ for $E_0 \gtrsim 2 \, {\rm keV}$ but also to measure its energy dependence with observatories such as  IXPE~\cite{ixpe}, eXTP~\cite{extp}, XL-Calibur~\cite{xcalibur}, and especially NGXP~\cite{ngxp} and XPP~\cite{xpp} at the highest energies.

In Fig.~\ref{perseusAllPolKeV} we have exhibited a particular realization of the photon-ALP beam propagation process, which depends on the particular choice of the orientation and coherence length of ${\bf B}^{\rm clu}$ and ${\bf B}_{\rm ext}$. The exact behavior of the latter quantities is unknown but only their statistical properties are known: thus, the propagation of the photon-ALP beam becomes a stochastic process. Therefore, by computing several realizations of the photon-ALP beam propagation, we can infer its statistical properties. While only one realization can be experienced by the photon-ALP beam at once and thus it represents the only physical possibility, the study of several realizations gives us information on the robustness of our results, as we vary the magnetic properties of the media within reasonable limits. Therefore, in Fig.~\ref{perseusDensKeV} we show the probability density function $f_{\Pi}$ for the final $\Pi_L$ of all realizations. We consider two benchmark energies $E_0 = 1 \, \rm keV$ and $E_0 = 10 \, \rm keV$. Because the final value $\Pi_L = 0$ is never the most probable result in Fig.~\ref{perseusDensKeV}, we conclude that our previous discussion about a possible detection of $\Pi_L > 0$ is robust.

In the HE band ($10^{-1}\, {\rm MeV} -10^4 \, \rm MeV$) we proceed by following the same strategy described above. In Fig.~\ref{perseusAllPolMeV} we show $P_{\gamma \to \gamma}$ -- and the corresponding $\Pi_L$ and $\chi$ -- for a specific realization of the photon-ALP beam propagation process, while we exhibit the statistical properties of the system in Fig.~\ref{perseusDensMeV} by plotting $f_{\Pi}$. From Fig.~\ref{perseusAllPolMeV} we see that the photon-ALP system is in the strong-mixing regime for almost all the HE range, which means that the plasma term and the mass term are now negligible with respect to the photon-ALP mixing term (see also~\cite{grSM}).  The binned data in Fig.~\ref{perseusAllPolMeV} show that the stability of the $\Pi_L$ value with respect to the energy in the HE range makes its detectability easier than in the X-ray band with observatories such as COSI~\cite{cosi}, e-ASTROGAM~\cite{eastrogam1,eastrogam2} and Amego~\cite{amego}. From Fig.~\ref{perseusDensMeV} we conclude that for the two benchmark energies $E_0 = 300 \, \rm keV$ and $E_0 = 3 \, \rm MeV$ the final value $\Pi_L = 0$ is never the most probable value. In particular, the case $E_0 = 3 \, \rm MeV$ shows that the most probable value for $\Pi_L$ in the strong-mixing regime is $\Pi_L \gtrsim 0.8$, which confirms the robustness of the previous discussion. 

We now move to the case $m_a = 10^{-10} \, \rm eV$. In the X-ray band the ALP mass term strongly dominates over the mixing term, so that photon-ALP conversion is totally inefficient and negligible. As a result $P_{\gamma \to \gamma}$ and the corresponding $\Pi_L$ are not modified by the photon-ALP interaction in the present situation. 

Instead, in the HE range the photon-ALP system turns out to be in the weak-mixing regime for $m_a = 10^{-10} \, \rm eV$, and the situation is totally similar to the previous case $m_a \lesssim 10^{-14} \, \rm eV$ in the X-ray band. Accordingly, we note from Fig.~\ref{perseusAllPolMeV-10} that the photon-ALP conversion is efficient for $E_0 \gtrsim 300 \, \rm keV$ and $\Pi_L$ increases from the initial value $\Pi_{L,0}=0$. Both $P_{\gamma \to \gamma}$ and $\Pi_L$ show an energy oscillatory behavior in almost all the considered range. As discussed in~\cite{galantiPol}, the extremely wide energy range, where the system is in the weak-mixing is due to the large variability of the properties of the media wherein the photon-ALP beam propagates (galaxy cluster, extragalactic space, Milky Way). The strong energy dependence is confirmed by the behavior of $\chi$. The binned data in Fig.~\ref{perseusAllPolMeV-10} suggest that we can expect a detectability of the above features for $E_0 \gtrsim 3 \, \rm MeV$ with observatories like COSI~\cite{cosi}, e-ASTROGAM~\cite{eastrogam1,eastrogam2} and AMEGO~\cite{amego}.

Figure~\ref{perseusDensMeV-10} confirms that the best energy range where to search for ALP-induced effects on $\Pi_L$ is $E_0 \gtrsim 3 \, \rm MeV$. For lower energies $f_{\Pi}$ shows that the most probable value for $\Pi_L$ is $\Pi_L \lesssim 0.2$, while at $E_0 = 3 \, \rm MeV$ it reads $\Pi_L \sim 0.3$.

\subsection{Coma cluster}

Coma is a nCC rich regular cluster, located at a redshift $z=0.0234$. It is a source of strong X-ray emission. We describe the Coma electron number density $n_e^{\rm clu}$ with the model~\cite{neComa}
\begin{equation}
\label{eq2c}
n_e^{\rm clu}(y)=n_{e,0}^{\rm clu} \left( 1+\frac{y^2}{r_{\rm core}^2} \right)^{-\frac{3}{2}\beta_{\rm clu}}~,
\end{equation}
where $n_{e,0}^{\rm clu}=3.44 \times 10^{-3} \, \rm cm^{-3}$, $r_{\rm core}=291 \, \rm kpc$ and $\beta_{\rm clu}=0.75$. Regarding the Coma magnetic field, we employ the model derived in~\cite{cluFeretti}. Correspondingly, we take the best fit values $B_0^{\rm clu} = 4.7 \, \mu {\rm G}$ and $\eta_{\rm clu}=0.5$ entering Eq.~(\ref{eq1}). Concerning the stochastic properties of ${\bf B}^{\rm clu}$, we take $q=-11/3$, $\Lambda_{\rm min}= 2 \, \rm kpc$ and $\Lambda_{\rm max}= 34 \, \rm kpc$ following~\cite{cluFeretti}.

As argued in Sec. III.A, photons in the central region of a galaxy cluster are expected to be emitted unpolarized both in the X-ray and in the HE band. For this reason, we take an initial degree of linear polarization $\Pi_{L,0} = 0$.

What we have discussed for the Perseus cluster in the X-ray and in the HE band -- and in the cases $m_a \lesssim 10^{-14} \, \rm eV$ and $m_a = 10^{-10} \, \rm eV$ -- is not very different from what happens for the Coma cluster. This is the reason why we prefer to highlight the few differences in the results for the two clusters instead of discussing Coma separately.

\begin{figure}
\centering
\includegraphics[width=0.5\textwidth]{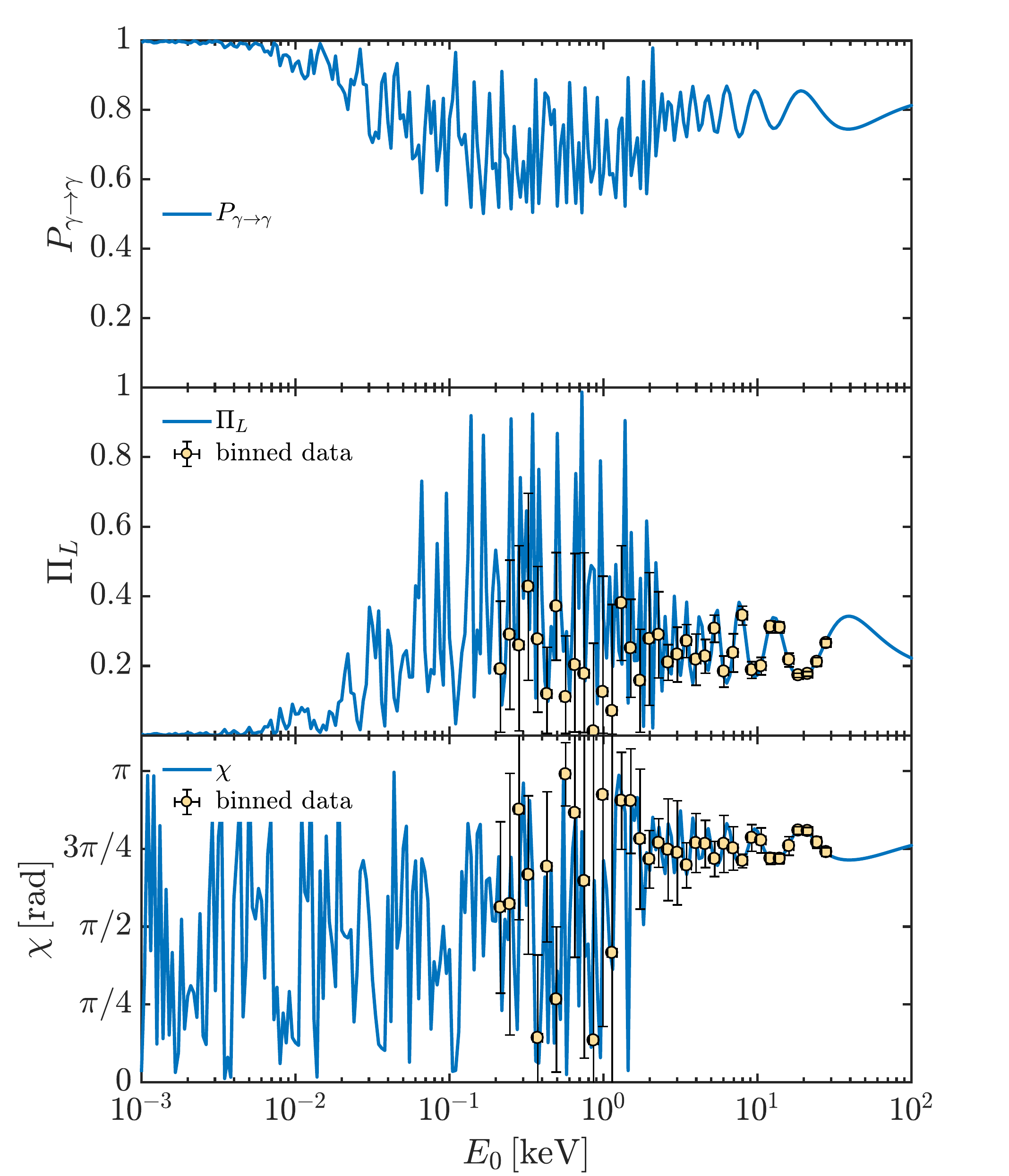}
\caption{\label{comaAllPolKeV} Coma cluster: photon survival probability $P_{\gamma \to \gamma }$ (upper panel), corresponding final degree of linear polarization $\Pi_L$ (central panel) and final polarization angle $\chi$ (lower panel) in the energy range $(10^{-3} -10^2)  \, {\rm keV}$. We take $g_{a\gamma\gamma}=0.5 \times 10^{-11} \, {\rm GeV}^{-1}$, $m_a \lesssim 10^{-14} \, {\rm eV}$. The initial degree of linear polarization is $\Pi_{L,0}=0$.}
\end{figure}

\begin{figure}
\centering
\includegraphics[width=0.5\textwidth]{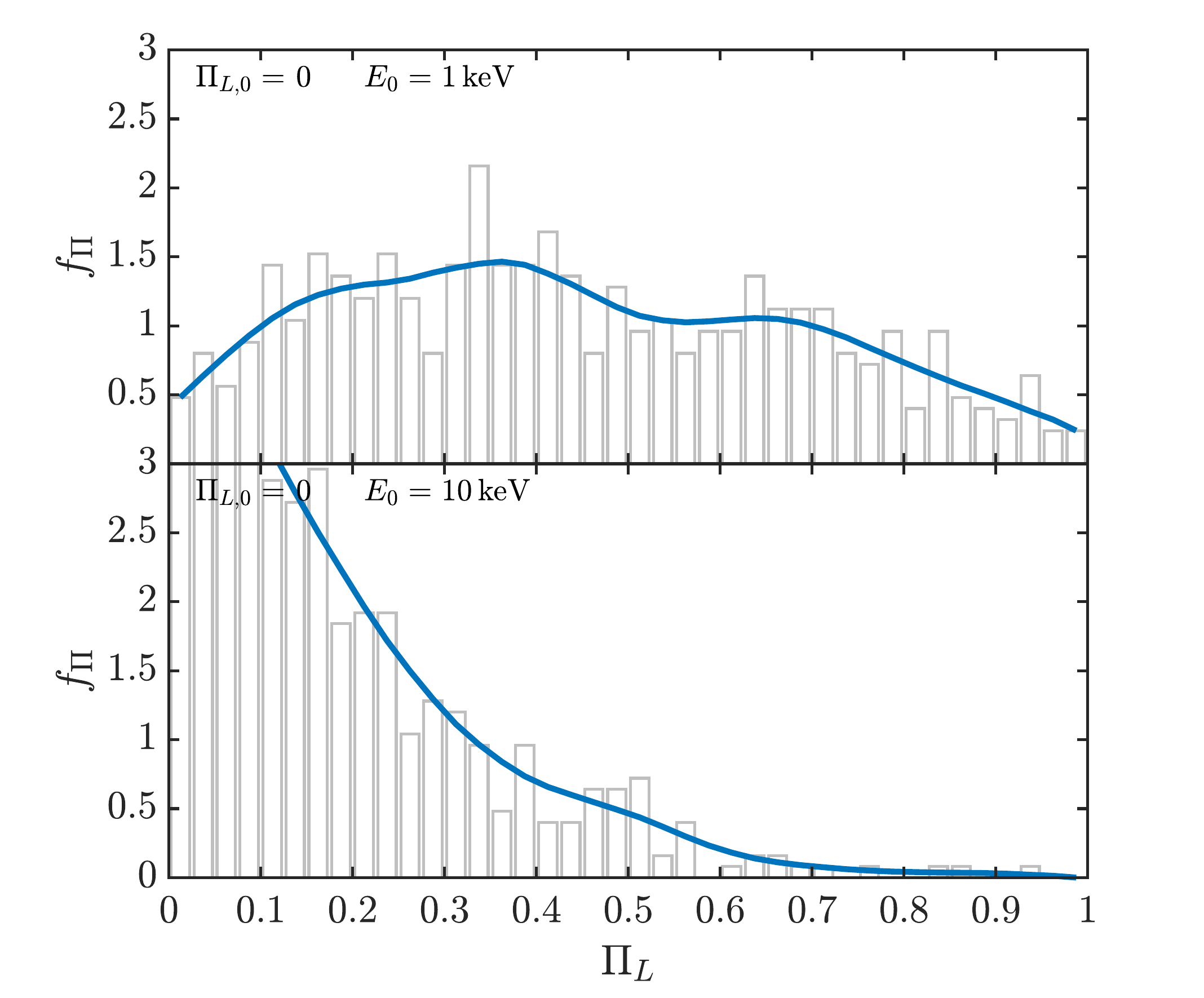}
\caption{\label{comaDensKeV} Coma cluster: probability density function $f_{\Pi}$ arising from the plotted histogram for the final degree of linear polarization $\Pi_L$ at $1 \, \rm keV$ (upper panel) and $10 \, \rm keV$ (lower panel) by considering the system in Fig.~\ref{comaAllPolKeV}. The initial photon degree of linear polarization is $\Pi_{L,0}=0$.}
\end{figure}

\begin{figure}
\centering
\includegraphics[width=0.5\textwidth]{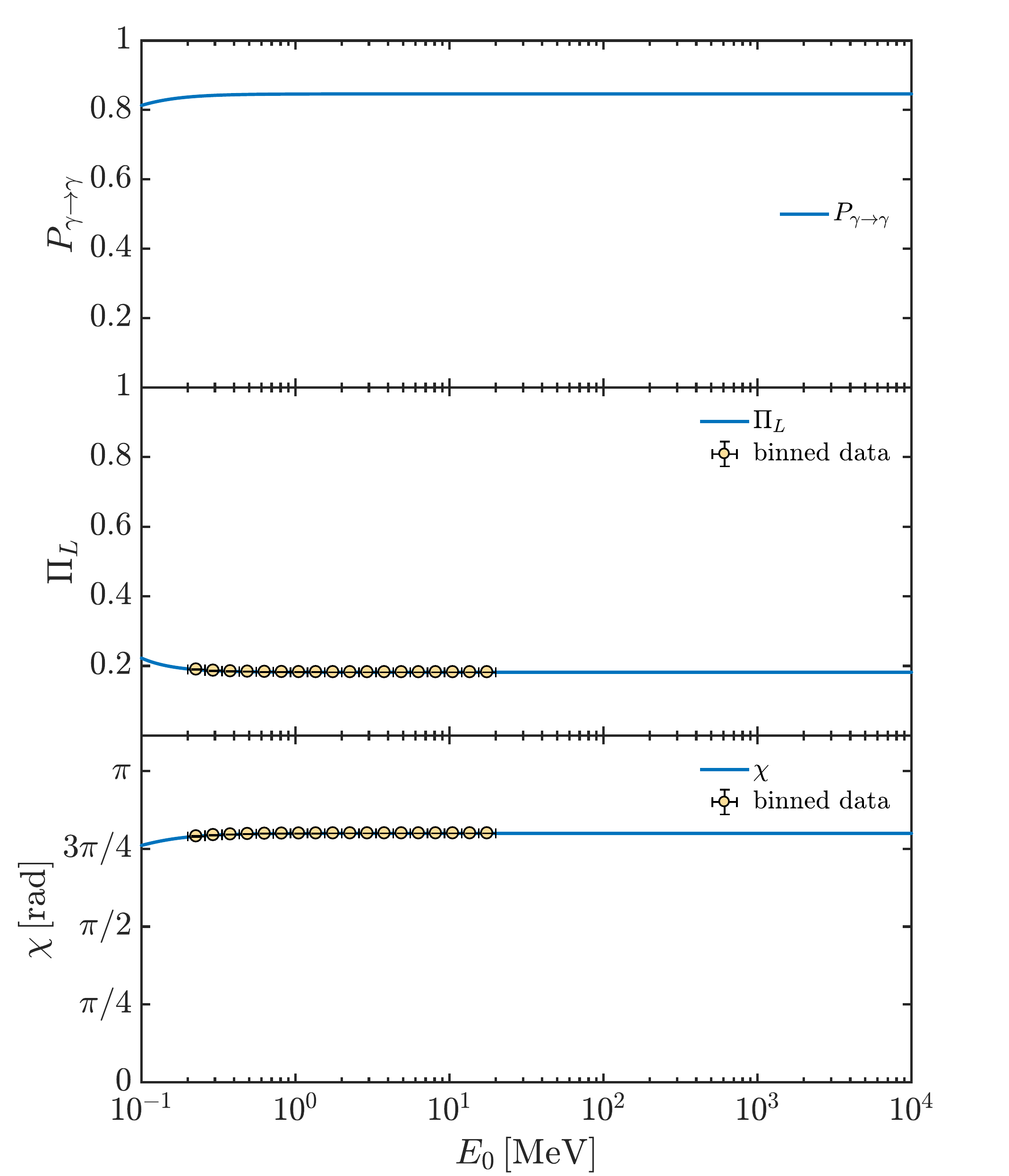}
\caption{\label{comaAllPolMeV} Coma cluster: same as Fig.~\ref{comaAllPolKeV} but in the energy range $(10^{-1}-10^4) \, \rm MeV$. We take $g_{a\gamma\gamma}=0.5 \times 10^{-11} \, \rm GeV^{-1}$, $m_a \lesssim 10^{-14} \, \rm eV$. The initial degree of linear polarization is $\Pi_{L,0}=0$.}
\end{figure}

\begin{figure}
\centering
\includegraphics[width=0.5\textwidth]{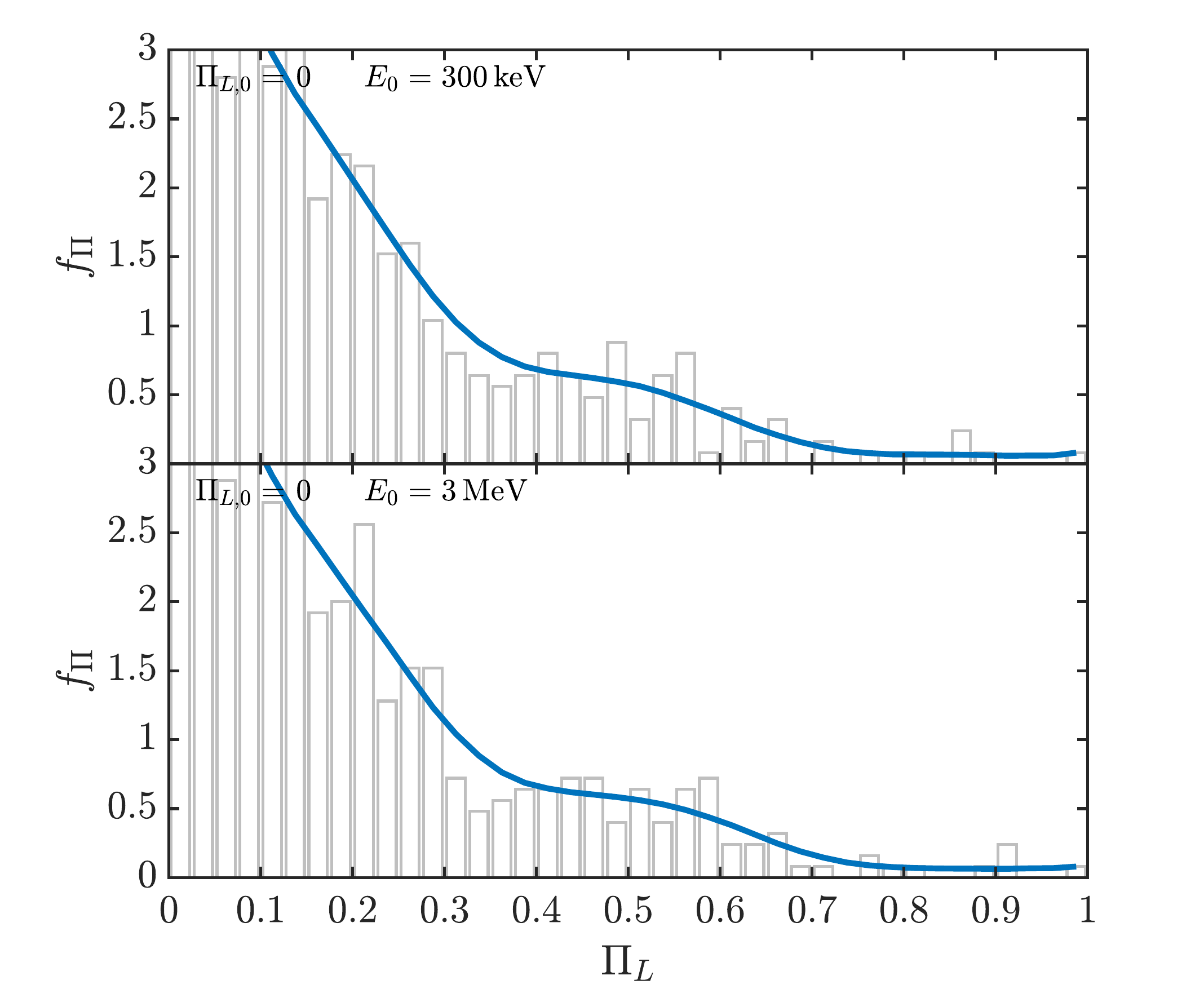}
\caption{\label{comaDensMeV} Coma cluster: same as Fig.~\ref{comaDensKeV} but for the energies $300 \, \rm keV$ (upper panel) and $3 \, \rm MeV$ (lower panel) by considering the system in Fig.~\ref{comaAllPolMeV}. The initial photon degree of linear polarization is $\Pi_{L,0}=0$.}
\end{figure}

\begin{figure}
\centering
\includegraphics[width=0.5\textwidth]{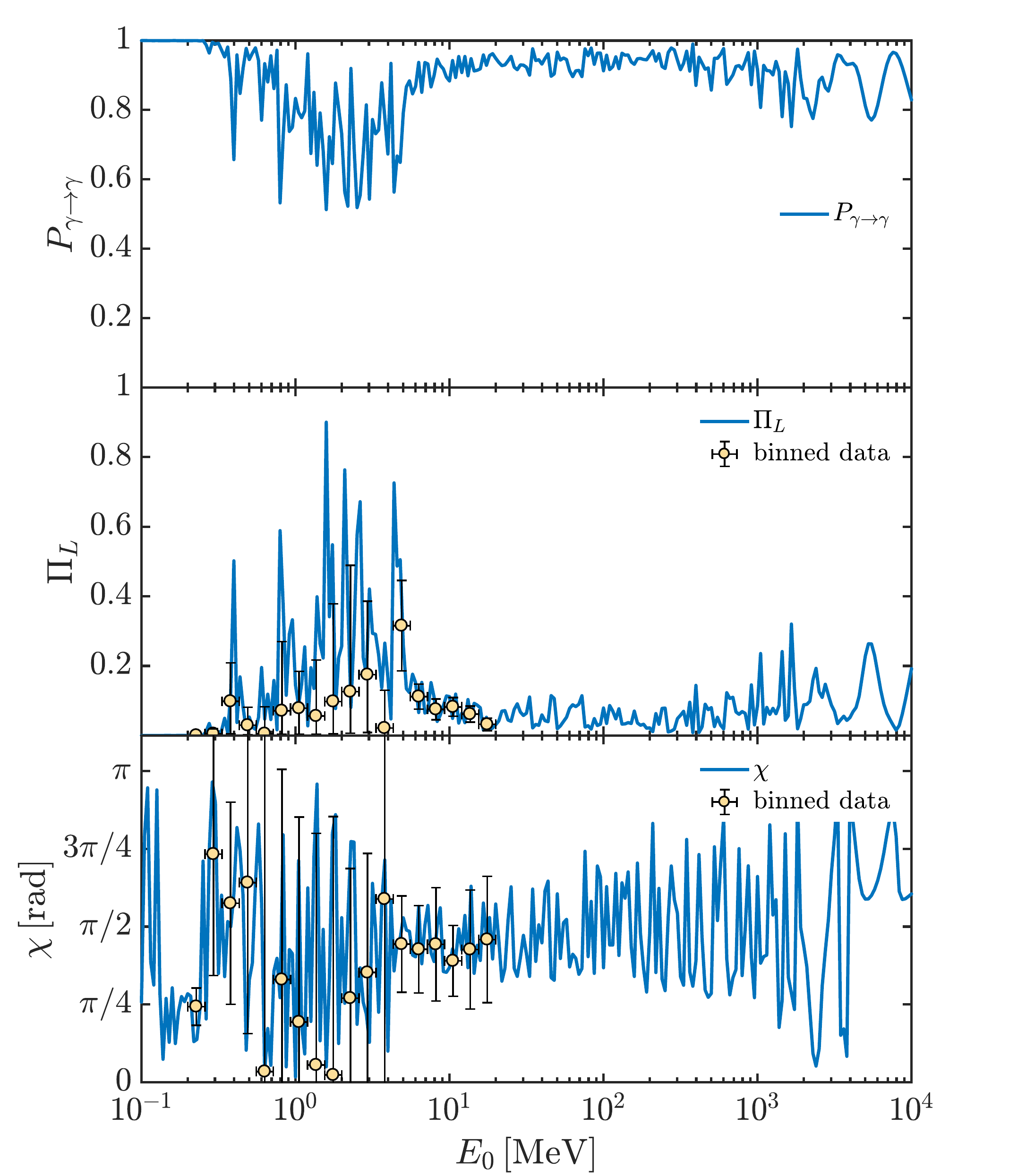}
\caption{\label{comaAllPolMeV-10} Coma cluster: same as Fig.~\ref{comaAllPolMeV}. We take $g_{a\gamma\gamma}=0.5 \times 10^{-11} \, \rm GeV^{-1}$, $m_a = 10^{-10} \, \rm eV$. The initial degree of linear polarization is $\Pi_{L,0}=0$.}
\end{figure}

\begin{figure}
\centering
\includegraphics[width=0.5\textwidth]{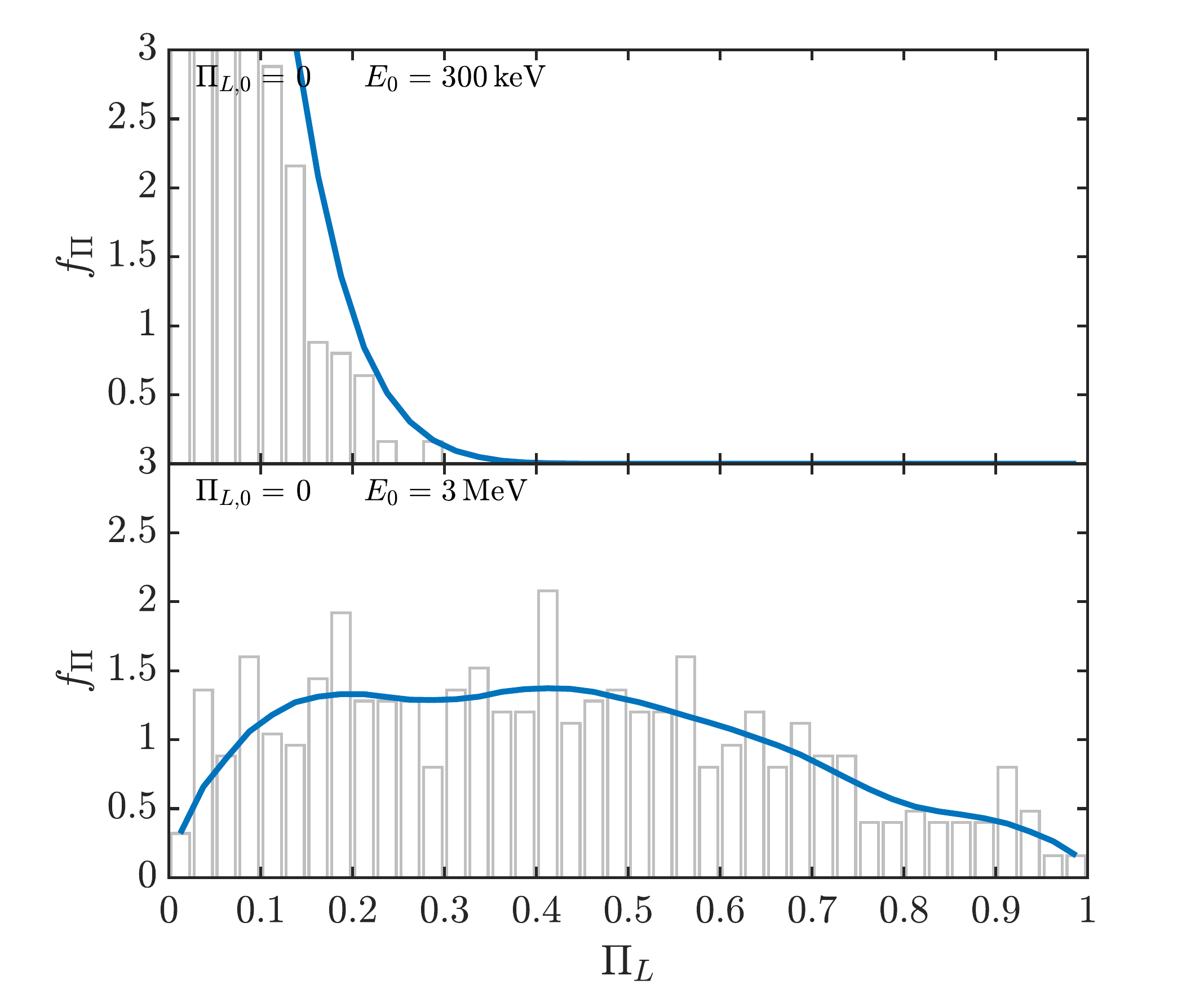}
\caption{\label{comaDensMeV-10} Coma cluster: same as Fig.~\ref{comaDensMeV} by considering the system in Fig.~\ref{comaAllPolMeV-10}. The initial photon degree of linear polarization is $\Pi_{L,0}=0$.}
\end{figure}

We start with the case $m_a \lesssim 10^{-14} \, \rm eV$. For a particular realization of the photon-ALP beam propagation in the UV-X-ray band ($10^{-3}\, {\rm keV} -10^2 \, \rm keV$), we report $P_{\gamma \to \gamma}$, $\Pi_L$ and $\chi$ in Fig.~\ref{comaAllPolKeV}, while the probability density function $f_{\Pi}$ associated with $\Pi_L$ for several realizations is reported in Fig.~\ref{comaDensKeV}. Note that Fig.~\ref{comaAllPolKeV} and  Fig.~\ref{perseusAllPolKeV} are qualitatively similar: the photon-ALP beam produced in the central region of the Coma cluster is in the weak-mixing regime in the energy range $(10^{-2} - 10^2) \, {\rm keV}$ for the same reasons explained for Perseus, while photon-ALP conversion is negligible for $E_0 \lesssim 10^{-2} \, \rm keV$. At higher energies, we find that $\Pi_L$ increases from the initial value $\Pi_{L,0}=0$ and shows an energy oscillatory behavior, as expected from the Perseus case. A firm detectability of $\Pi_L>0$ is expected for $E_0 \gtrsim 2 \, \rm keV$ and also its energy dependence seems observable with IXPE~\cite{ixpe}, eXTP~\cite{extp}, XL-Calibur~\cite{xcalibur}, and especially with NGXP~\cite{ngxp} and XPP~\cite{xpp} at the highest energies.

Figure~\ref{comaDensKeV} shows that for lower energies (see the upper panel where $E_0 = 1 \, \rm keV$) the most probable value of $\Pi_L$ is $\Pi_L \sim 0.3$, while at higher energies (see the lower panel where $E_0 = 10 \, \rm keV$) the most probable result turns out to be $\Pi_L < 0.2$. The latter fact differentiates Coma from Perseus. Thus, for Coma the best energies to search for signals of $\Pi_L >0$ appear to be in the range $2 \, {\rm keV} \lesssim E_0 \lesssim 10 \, \rm keV$. The reason behind the different behavior of the two clusters lies in the different strength of $B_0^{\rm clu}$. In particular, the behavior of $f_{\Pi}$ is similar at lower energies ($1 \, \rm keV$, see upper panels of Figs.~\ref{perseusDensKeV} and~\ref{comaDensKeV}), since the photon-ALP beam propagates well inside the weak mixing regime (see also Figs.~\ref{perseusAllPolKeV} and~\ref{comaAllPolKeV}, respectively) and the energy dependence allows to $\Pi_L$ to assume a wide range of values. Instead, with the photon-ALP system approaching to the strong mixing regime at higher energies ($10 \, \rm keV$, see lower panels of Figs.~\ref{perseusDensKeV} and~\ref{comaDensKeV}), the higher strength of $B_0^{\rm clu}$ for Perseus produces a larger modification in the final values of $\Pi_L$, while for Coma the final $\Pi_L$ generally settles down to lower values because of the smaller $B_0^{\rm clu}$ (see also Figs.~\ref{perseusAllPolKeV} and~\ref{comaAllPolKeV}, respectively).

From Fig.~\ref{comaAllPolMeV} we find that the photon-ALP beam coming from Coma propagates in the strong-mixing regime in the HE band ($10^{-1}\, {\rm MeV}-10^4 \, \rm MeV$), as it happens for Perseus. The stability of the binned data concerning the $\Pi_L$ value -- as the energy increases -- suggests that observatories like COSI~\cite{cosi}, e-ASTROGAM~\cite{eastrogam1,eastrogam2} and AMEGO~\cite{amego} may detect this signal.

However, $f_{\Pi}$ in Fig.~\ref{comaDensMeV} shows that in the strong-mixing regime the most probable value for $\Pi_L$ is $\Pi_L \lesssim 0.4$, so that even if a signal is surely detectable the configuration of ${\bf B}^{\rm clu}$ and ${\bf B}_{\rm ext}$ should be favorable in order to produce $\Pi_L$ sensibly larger than zero. The reason for the different behavior of $f_{\Pi}$ concerning Perseus and Coma (see Figs.~\ref{perseusDensMeV} and~\ref{comaDensMeV}, respectively) is the same expressed for the X-ray band: in the strong mixing regime the higher strength of $B_0^{\rm clu}$ of Perseus is more efficient in producing a larger modification to the final $\Pi_L$ than in the case of Coma.

We now turn our attention to the case $m_a = 10^{-10} \, \rm eV$. What we have discussed for Perseus in the X-ray band retains its validity for Coma: ALP effects are negligible. Hence, $P_{\gamma \to \gamma}$ and the corresponding $\Pi_L$ are unchanged.

In the HE range the behavior of Coma is similar to that of Perseus. In particular, Fig.~\ref{comaAllPolMeV-10} shows that the photon-ALP interaction starts to be efficient for $E_0 \gtrsim 300 \, \rm keV$ and the photon-ALP beam propagates in the weak-mixing regime in almost all the energy range under consideration for the same reasons discussed for Perseus. As a consequence, $P_{\gamma \to \gamma}$ and the corresponding $\Pi_L$ show an energy oscillatory behavior. From the binned data in Fig.~\ref{comaAllPolMeV-10} we expect that observatories like COSI~\cite{cosi}, e-ASTROGAM~\cite{eastrogam1,eastrogam2} and AMEGO can detect ALP features for $E_0 \gtrsim 3 \, \rm MeV$. Again, we find the same conclusions derived for Perseus.

The analysis of $f_{\Pi}$ in Fig.~\ref{comaDensMeV-10} demonstrates that $E_0 \gtrsim 3 \, \rm MeV$ is the best energy range where to search for ALP effects on photon polarization, since the most probable value for $\Pi_L$ at $E_0 = 3 \, \rm MeV$ is around 0.3, while for lower energies (see the case $E_0 = 300 \, \rm keV$ in the upper panel of Fig.~\ref{comaDensMeV-10}) the most probable value for $\Pi_L$ is smaller than 0.2. The latter findings are still in agreement with the results derived for the Perseus cluster. At lower energies ($300 \, \rm keV$, see upper panels of Figs.~\ref{perseusDensMeV-10} and~\ref{comaDensMeV-10}) the photon-ALP conversion is quite inefficient for both Perseus and Coma (see also Figs.~\ref{perseusAllPolMeV-10} and~\ref{comaAllPolMeV-10}, respectively) producing in either case a small modification in the values of the final $\Pi_L$. But at higher energies ($3 \, \rm MeV$, see lower panels of Figs.~\ref{perseusDensMeV-10} and~\ref{comaDensMeV-10}) the photon-ALP beam propagates more efficiently inside the weak mixing regime for both Perseus and Coma (see also Figs.~\ref{perseusAllPolMeV-10} and~\ref{comaAllPolMeV-10}, respectively), so that the final $\Pi_L$ gets effectively more modified for both the clusters.

\subsection{Polarization detectability}

Both the Perseus and Coma clusters represent very promising targets for polarization studies both in the X-ray and in the HE band as far as the detectability of the features produced by the photon-ALP interaction is concerned. These effects are primarily produced by photon-ALP interaction inside the cluster, while its contribution in the other crossed regions is less important. In particular, although a rather high extragalactic magnetic field strength $B_{\rm ext}=1 \, \rm nG$ producing an effective photon-ALP conversion is the most probable scenario (see Sec. III.B), we have checked that an inefficient photon-ALP conversion in the extragalactic space -- arising by taking $B_{\rm ext}< 10^{-15} \, \rm G$ -- does not substantially affect our previous results. We find  only a slight and negligible dimming of the broadening of $f_{\Pi}$ in all the previous figures. At low energies, where the photon-ALP system lies in the weak-mixing regime, the ALP-induced oscillations of $\Pi_L$ and $\chi$ with respect to the energy are very quick, as shown by Fig.~\ref{perseusAllPolKeV} for Perseus and by Fig.~\ref{comaAllPolKeV} for Coma especially in the $(0.1-1) \, \rm keV$ decade. Since within a single bin we expect many oscillations with respect to the energy in the latter situation, the high dispersion in the values assumed by $\Pi_L$ and $\chi$ provokes a larger error bar in the binned data at lower energies. The latter effect decreases as the energy increases until it becomes negligible in the strong-mixing regime, as shown by Fig.~\ref{perseusAllPolMeV} for Perseus and by Fig.~\ref{comaAllPolMeV} for Coma.

As already mentioned in Sec. II, the only firm constraint about the ALP parameter space derives from the CAST experiment~\cite{cast}. However, new bounds about the photon-ALP system parameters ($m_a, g_{a\gamma\gamma}$) have recently appeared in the literature and suggest us a preference for a specific model~\cite{limFabian,limJulia,limKripp,limRey2}. In particular, the case $[m_a \lesssim 10^{-14} \, {\rm eV} , g_{a\gamma\gamma} = 0.5 \times 10^{-11} \, {\rm GeV^{-1}}]$ is disfavored by~\cite{limFabian,limJulia,limKripp,limRey2} with respect to the other considered in this paper $[m_a = 10^{-10} \, {\rm eV}, g_{a\gamma\gamma} = 0.5 \times 10^{-11} \, {\rm GeV}^{-1}]$, which is within all current bounds. Therefore, while we cannot exclude the case $[m_a \lesssim 10^{-14} \, {\rm eV}, g_{a\gamma\gamma} = 0.5 \times 10^{-11} \, {\rm GeV}^{-1}]$, we consider the case $[m_a = 10^{-10} \, {\rm eV}, g_{a\gamma\gamma} = 0.5 \times 10^{-11} \, {\rm GeV}^{-1}]$ as more probable. In addition, note that the case $[m_a = 10^{-10} \, {\rm eV}, g_{a\gamma\gamma} = 0.5 \times 10^{-11} \, {\rm GeV}^{-1}]$ is compatible with the two hints at ALP existence~\cite{trgb2012,grdb} and with the explanation of the GRB 221009A detection at $18 \, \rm TeV$ by LHAASO and at $251 \, \rm TeV$ by Carpet-2~\cite{grtGRB}. As a result, since the case $[m_a = 10^{-10} \, {\rm eV},g_{a\gamma\gamma} = 0.5 \times 10^{-11} \, {\rm GeV^{-1}}]$ produces polarization effects in the HE range but not in the X-ray band, we conclude that the best observatories that can detect ALP-induced polarization effects in galaxy clusters are COSI~\cite{cosi}, e-ASTROGAM~\cite{eastrogam1,eastrogam2} and AMEGO~\cite{amego}. Still, we cannot exclude a detection in the X-ray band similar to what we have reported in the figures above. Obviously, many other possibilities remain to be explored. An ALP with a mass $ m_a = 2 \times 10^{-12} \, \rm eV$ and the same coupling $g_{a\gamma\gamma}= 0.5 \times 10^{-11} \, \rm GeV^{-1}$ -- which is within the above-mentioned bounds~\cite{limFabian,limJulia,limKripp,limRey2} -- would exhibit an intermediate behavior between the two situations considered above, namely $m_a \lesssim 10^{-14} \, \rm eV$ and $m_a = 10^{-10} \, \rm eV$. In particular, both the region where photon-ALP conversion is efficient and the strong-mixing regime would start at higher energies with respect to the case $[m_a \lesssim 10^{-14} \, {\rm eV}, g_{a\gamma\gamma} = 0.5 \times 10^{-11} \, {\rm GeV}^{-1}]$  and lower with respect to the case $[m_a = 10^{-10} \, {\rm eV}, g_{a\gamma\gamma} = 0.5 \times 10^{-11} \, {\rm GeV}^{-1}]$.

\section{Conclusions}

In this paper, we have investigated the effects of the photon-ALP interaction on both the final degree of linear polarization $\Pi_L$ and the polarization angle $\chi$ of photons produced in the central region of two regular clusters. We have chosen Perseus and Coma, and we have addressed both the X-ray and the HE bands. In either case, photons are expected to be emitted as unpolarized ($\Pi_{L,0}=0$). Our findings are consistent with those obtained in~\cite{galantiPol} for generic galaxy clusters. We have employed the state-of-the-art knowledge about the astrophysical media (galaxy cluster, extragalactic space, Milky Way) crossed by the photon-ALP beam. We have considered ALP parameters within the firm bound derived by CAST~\cite{cast}, we have taken $g_{a\gamma\gamma}=0.5 \times 10^{-11} \, \rm GeV^{-1}$ and the two values for the ALP mass: (i) $m_a \lesssim 10^{-14} \, \rm eV$, (ii) $m_a = 10^{-10} \, \rm eV$. We have found features in the final $\Pi_L$ induced by the photon-ALP interaction and we have performed a first estimate of their detectability with observatories like IXPE~\cite{ixpe}, eXTP~\cite{extp}, XL-Calibur~\cite{xcalibur}, NGXP~\cite{ngxp} and XPP~\cite{xpp} in the X-ray band, and COSI~\cite{cosi}, e-ASTROGAM~\cite{eastrogam1,eastrogam2} and AMEGO~\cite{amego} in the HE band. We have also investigated the probability density function $f_{\Pi}$ of $\Pi_L$ associated with many realizations of the photon-ALP beam propagation process. Our conclusions can be summarized as follows.

\begin{enumerate}[(i)]

\item In the X-ray band only the case $m_a \lesssim 10^{-14} \, {\rm eV}$ produces observable effects on $\Pi_L$. Instead, for $m_a = 10^{-10} \, {\rm eV}$ the ALP mass  effect is so large that photon-ALP conversion is totally negligible and gives rise to no features on the final $\Pi_L$. Results for Perseus and Coma are qualitatively similar. In the case $m_a \lesssim 10^{-14} \, \rm eV$ the photon-ALP beam propagates in the 
weak-mixing regime so that the final $\Pi_L$ shows an energy-dependent behavior. From an estimate of the energy resolution of observatories like IXPE~\cite{ixpe}, eXTP~\cite{extp}, XL-Calibur~\cite{xcalibur}, NGXP~\cite{ngxp} and XPP~\cite{xpp}, we expect that an ALP-induced signal about $\Pi_L>0$ can be detected for $E_0 \gtrsim 2 \, \rm keV$. The analysis of $f_{\Pi}$ -- associated with many realizations of the photon-ALP beam propagation -- shows that we can expect a more probable signal of $\Pi_L >0$ for $E_0$ around $10 \, \rm keV$ for Perseus rather than for Coma. Instead, for both clusters $\Pi_L >0$ is the most likely result at lower energies. Thus, the detection of a possible signal of $\Pi_L>0$ appears as rather robust.

\item In the HE range both the case $m_a \lesssim 10^{-14} \, \rm eV$ and $m_a = 10^{-10} \, \rm eV$ lead to an efficient photon-ALP conversion and features in $\Pi_L$. Also in the HE range the behavior of Perseus and Coma are similar. In the case $m_a \lesssim 10^{-14} \, \rm eV$ the photon-ALP beam propagates in the strong-mixing regime and the behavior of $\Pi_L$ turns out to be energy independent. For this reason, a signal with $\Pi_L>0$ can very well be observed by detectors like COSI~\cite{cosi}, e-ASTROGAM~\cite{eastrogam1,eastrogam2} and AMEGO~\cite{amego}. The probability of getting a signal with $\Pi_L>0$ is greater for Perseus than for Coma, as shown by the study of $f_{\Pi}$. In the case $m_a = 10^{-10} \, \rm eV$ the qualitative behaviors of Perseus and Coma are almost identical. The photon-ALP beam propagates in the weak-mixing regime and $\Pi_L$ shows an energy oscillatory behavior. The analysis of both the binned data about $\Pi_L$ and of $f_{\Pi}$ suggests that the range $E_0 \gtrsim 3 \, \rm MeV$ represents the best energy region where to expect the highest probability of measuring a signal with $\Pi_L >0$ from observatories such as COSI~\cite{cosi}, e-ASTROGAM~\cite{eastrogam1,eastrogam2} and AMEGO~\cite{amego}.

\end{enumerate}

As discussed above, we cannot exclude a detection of ALP-induced polarization effects in the X-ray band since the only firm bound on ALP parameters is that from CAST~\cite{cast}, but we believe the HE band to be the best window to search for these effects. 

In conclusion, both Perseus and Coma represent good targets for the study of ALP-induced effects on $\Pi_L$ with some preference for Perseus. Instead, for a detailed study of the behavior of $\Pi_L$ as the energy varies, Coma appears as a slightly better candidate.

Different physically consistent models concerning the behavior of ${\bf B}^{\rm clu}$ and $n_e^{\rm clu}$ do not produce a strong modification of our final results, as already noted in~\cite{galantiPol}. A higher impact would be produced by a large modification of the strength of $B_0^{\rm clu}$ and of the central value of $n_e^{\rm clu}$, but these quantities are reasonably quite well known for both Perseus and Coma. 

Since the ALP-induced polarization effects considered in the present paper can only increase the initial degree of linear polarization $\Pi_{L,0}=0$, the same features cannot instead be produced by Lorentz invariance violation (LIV), since the LIV trend is to reduce $\Pi_L$~\cite{LIVpol}.

ALPs with the properties investigated in this paper can also be detected by the new generation of VHE gamma-ray observatories like CTA~\cite{cta}, HAWC~\cite{hawc}, GAMMA-400~\cite{g400}, LHAASO~\cite{lhaaso}, TAIGA-HiSCORE~\cite{desy} and HERD~\cite{herd}. In addition, these ALPs can be directly detected by laboratory experiments like the upgrade of ALPS II at DESY~\cite{alps2}, the planned IAXO~\cite{iaxo,iaxo2} and STAX~\cite{stax}, and with the techniques developed by Avignone and collaborators~\cite{avignone1,avignone2,avignone3}. Moreover, if ALPs turn out to constitute of the bulk of the dark matter, then they can be detected also by the planned ABRACADABRA experiment~\cite{abracadabra}.

\section*{Acknowledgments}

G.G. acknowledges a contribution from the grant ASI-INAF 2015-023-R.1. M.R. acknowledges the financial support by the TAsP grant of INFN. This work was made possible also by the funding of the INAF Mini Grant `High-energy astrophysics and axion-like particles', PI: Giorgio Galanti.


\begin{thebibliography}{20}
\bibitem{noteCluPol} As already discussed in~\cite{galantiPol}, an initial $\Pi_{L,0}$ slightly bigger than zero in the HE band ($\Pi_{L,0} = 0$ however represents the most probable situation) does not substantially change the final results especially in the cases where the photon-ALP interaction produces strong polarization effects, which can be observed. This is the reason why we consider the most probable case $\Pi_{L,0} = 0$ only.
\bibitem{noteDetectability} If we adopt the polarimetric conventions, the detectability level requirement is 99$\%$.
\bibitem{noteNGC1275} The Perseus cluster hosts the active galaxy NGC~1275, which represents an identified source of photons both in the X-ray and HE band. Yet, because of the angular extension of Perseus, with IXPE, and with the other proposed missions of X-ray polarimetry, it is possible to separate the point-like emission of NGC~1275 and of surrounding filaments and cavity from the diffuse emission from the cluster, which is of our interest in the present analysis.
\end{thebibliography}
\end{document}